\documentclass{aastex62}
\shorttitle{Causes of the Diurnal Variation observed in $\gamma$-ray Spectrum using NaI (Tl) Detector}
\shorttitle{Causes of the Diurnal Variation observed in $\gamma$-ray Spectrum using NaI (Tl) Detector}
\shortauthors{Datar et al.}

\begin{document}

\title{Causes of the Diurnal Variation observed in $\gamma$-ray Spectrum using NaI (Tl) Detector}

\correspondingauthor{Gauri Datar}
\email{datar.gouri@gmail.com, gdatar16@iigs.iigm.res.in}

\author[0000-0003-4971-6010]{Gauri Datar}
\affil{Indian Institute of Geomagnetism, Navi Mumbai, 410218, India}

\author{Geeta Vichare}
\affil{Indian Institute of Geomagnetism, Navi Mumbai, 410218, India}

\author{C. Selvaraj}
\affil{Equatorial Geophysical Research Laboratory, Tirunelveli, 627011, India}

\author{Ankush Bhaskar}
\affiliation{Heliophysics Science Division, NASA/Goddard Space Flight Center, Greenbelt, 20770, USA}
\affiliation{University Corporation for Atmospheric Research, Boulder, CO, 80307 USA}

\author{Anil Raghav}
\affiliation{University of Mumbai, Department of Physics, Santacruz (Mumbai), 400098, India}

\begin{abstract}

Present paper studies the $\gamma$-ray spectrum observed using NaI(Tl) scintillation detector over a year. The paper reports the presence of a distinct diurnal pattern in the total number of $\gamma$-ray counts detected by the NaI detector. The counts start decreasing after sunrise and show gradual recovery after sunset. The amplitude of this variation is quite significant ( $>$ 10\%) at the observation site Tirunelveli, South India. However, further investigation based on different energy ranges reveals that the mentioned diurnal pattern is actually present only in the energies related to the terrestrial background radioactivity. The study demonstrates that the pattern is associated with the radioactivity of isotopes of radon ($^{222}$Rn and $^{220}$Rn ) and their daughter radionuclides. The amplitude of the diurnal variation is found to have seasonal dependence, with the lowest amplitude during June--October ($\sim$ 2\%), and highest in April--May months ($\sim$ 14\%). The dependence of the amplitude of the diurnal variation on various atmospheric parameters, namely temperature, relative humidity, and vertical wind is examined. The observed diurnal pattern is attributed to the generation of the inversion layer. The distribution of concentration of radon and thoron progenies in an air column changes with the vertical mixing and atmospheric boundary layer (ABL), resulting in the diurnal variation of $\gamma$-rays.

\end{abstract}


\section{Introduction} \label{sec:intro}
%\linenumbers
The surrounding atmosphere contains abundant radiation, with a daily influx from the Sun, cosmic rays (CR), in addition to radiation of terrestrial origins. The near-Earth environment is abundant with a vast number of particles of various energies and a broad spectrum of radiation. One of the primary components of radiation affecting humans is $\gamma$-rays of varying energies. Due to very high penetration power, $\gamma$-rays can damage human organs and bone marrow, thus making their high exposure biologically hazardous. However, all of us are regularly exposed to a certain level of radiation, i.e., \lq{Background radiation}', that collectively represents all types of radiation together that are present in our surroundings at all times. The significant natural sources of $\gamma$-rays on Earth are (1) terrestrial radioactive decay, and (2) secondary particles generated during the interaction of CR with atmospheric nuclei \citep{united2000sources}. However, there are other natural sources as well that have a relatively lesser contribution, e.g., lightning, terrestrial $\gamma$-ray flashes (TGF) in thunderstorm produce $\gamma$-rays \citep{fishman1994discovery, briggs2013terrestrial}. Furthermore, fission in nuclear reactors, high energy physics experiments, nuclear explosions, and accidents are some of the artificial sources of $\gamma$-rays. 

Even though background radiation is present everywhere on Earth, the extent can differ from place to place, especially where human-made radiation can be in high concentration. Background radiation is measured in such areas to monitor the radiation dose received by the general population. The most significant radionuclides that enter the body are of terrestrial origin. Terrestrial radioactivity is present mainly due to radiations from certain radioactive isotopes present in the Earth's crust since its formation. Among the radioactive isotopes, thorium ($^{232}$Th), uranium ($^{235}$U, $^{238}$U), and the daughter radionuclides formed in their decay chains are the major contributors (the decay chains are shown in the Appendix). These isotopes and their daughter radionuclides emit $\gamma$-rays of different energies until a stable isotope is reached in the chain \citep{knoll2010radiation}. Among these radionuclides, radon gas and its decay products constitute the principal part of the dosage that we inhale constantly. Radon levels in a particular geographic region depend on the uranium and thorium content of the soil. The $^{222}$Rn decay products affecting environmental $\gamma$-ray intensity are $^{214}$Pb and $^{214}${Bi}. 

As $^{222}$Rn is chemically inert, and it has a well-defined source and sink \citep{jacobi1963vertical}; it is considered as a useful tracer for studying the vertical mixing of the pollutants in the atmospheric boundary layer (ABL) (e.g., \citet{liu1984radon, galeriu2011radon}). Atmospheric abundance of radon is connected with its exhalation rate from the soil, which is estimated by \citet{wilkening1972radon} as the mean flux of 0.75 atom cm$^{-2}$ sec$^{-1}$ with a range from 0.01 to 2.5 atoms cm$^{-2}$ sec$^{-1}$ or 0.72 to 1.2 atoms cm$^{-2}$ sec$^{-1}$ for ice-free terrains (e.g., \citet{griffiths2010map}). It is also affected by factors such as soil composition, moisture content, porosity, and permeability \citet{szegvary2009european, minty2004radon, manohar2013radon, arora2017assesment}. Besides, to understand how the radon concentration varies with height, the vertical profile of radon concentration has been studied for a range of altitudes (near-surface to stratosphere), and geographical locations (inland, coastal, islands, and open ocean), e.g., \citet{williams2011vertical} (and other studies enlisted therein in Table 1), \citet{baldoncini2017exploring}. 

Many studies have investigated the relationship between rainfall and $^{222}$Rn concentration \citep{takeuchi1982rainout,minato1983estimate,horng2003rainout,gusev2015simulation}. The diurnal variation in the concentration of radon and its relation with the metrological parameters of temperature, relative humidity, and pressure has been studied by many researchers such as \citet{el2001diurnal, schubert2002diurnal, desideri2006monitoring, galeriu2011radon, banjanac2012daily, VICTOR2019105118}.
Further, the variations of radon concentration in the atmosphere in different seasons have been studied, and it was found by \citet{tchorz2018variations} that the highest radon concentrations were during autumn and the lowest during winter. Besides the diurnal and seasonal variations in the atmospheric radon, there have been reports on the diurnal variations of the radon concentration in the upper soil layers and at the soil-air interface with meteorological parameters \citep{schubert2002diurnal, VICTOR2019105118}. 

As discussed earlier, the other significant source of $\gamma$-rays on Earth is cosmic rays in the atmosphere. CR continually arrive at Earth, and after entering the atmosphere, they tend to collide with atmospheric atoms and molecules. Further, the spallation process produces a variety of lighter particles, which are called secondary cosmic rays (SCR). In general, muons and neutrinos originate in the decay chain of charged mesons, while electrons and photons are produced as a result of the decays of neutral mesons \citep{patrignani2016review}. Photons are generated due to annihilation processes as well (e.g., e$^-$e$^+$, $\mu^-\mu^+$). Among SCR, the most abundant particles at sea level are muons, neutrons, neutrinos, photons. Muons with energies greater than 250 MeV are the most abundant secondary particles, while photons are predominant in the energy range between 100 keV to $\sim$10 MeV. 

\citet{chin1968solar} studied the variation of the $\gamma$-rays (photon component of SCR) in the energy range 3.8 -- 183 MeV using NaI (Tl) spectrometer. They observed the dependence of the time of maximum amplitude on energy using 16 energy bands in the range 3.8 -- 183 MeV. For energies $<$ 50 -- 60 MeV, \citet{chin1968solar} observed that the times of maximum intensity move towards later hours with increasing energy, and the amplitudes decrease with energy. In their other paper, using the same dataset, \citet{chin1968barometric} found that the photon intensity depends only on the atmospheric pressure, and does not vary significantly with the ground temperature. \citet{katase1982variation} reported the presence of a diurnal pattern in the intensity of $\gamma$-rays obtained using Ge (Li) spectrometer. Recently, using NaI(Tl) observations, \citet{vichare2018equatorial} described the diurnal variation of the $\gamma$-ray counts, and a similar pattern was also noticed by \citet{raghav2013confirmation, raghav2015low}. However, none of these have studied the source of this variation.  In fact, there are only a few studies to date, focusing on the diurnal pattern of the $\gamma$-ray spectrum. The present paper aims to conduct an in-depth analysis of the diurnal pattern observed in the $\gamma$-ray flux and the parameters responsible, using $\gamma$-ray data spanning one year from January 2017 to December 2017. Such long term data also facilitates examining the seasonal dependence, if any. 

\section{Data description} \label{sec:style}

\begin{figure*}
\vspace*{2mm}
\begin{center}
\includegraphics[width=8cm]{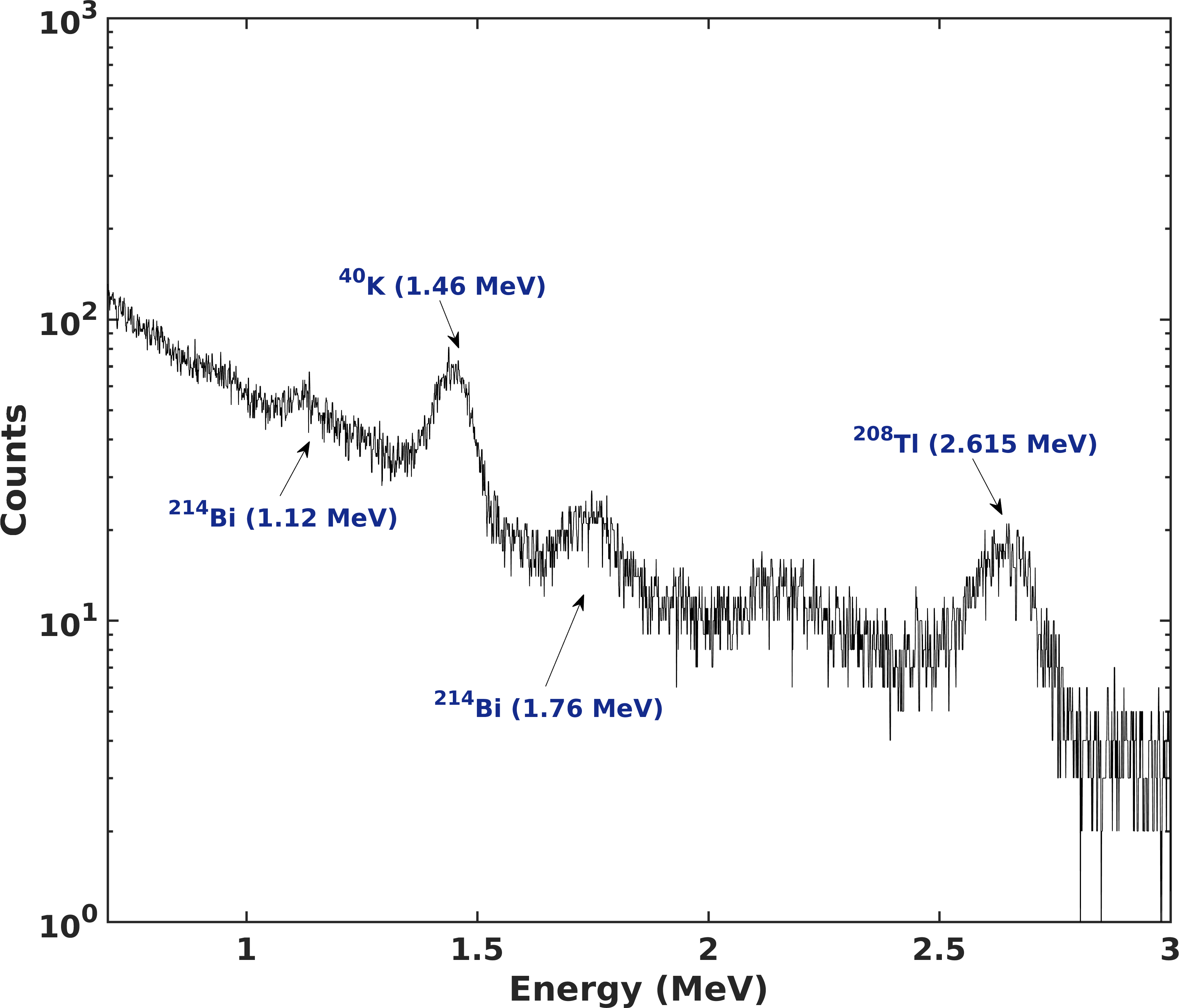} 
\end{center}
\caption{\textit{Background spectrum observed in NaI(Tl) data; x-axis shows energy in MeV, and the corresponding counts on y-axis are accumulated in one hour.}}
 \label{fig:one}
\end{figure*}

The $\gamma$-ray flux data used in the present analysis is obtained from  a rectangular cuboid sized NaI (Tl) scintillation detector, of dimension 10.16 cm $\times$ 10.16 cm $\times$ 40.64 cm (4" $\times$ 4" $\times$ 16"), installed at Tirunelveli (Geographic Coordinates: 8.71{$^\circ$}N, 77.76{$^\circ$}; Geomagnetic Coordinates: 0.28{$^\circ$}N, 151.01{$^\circ$}E) in South India.  The detector is surrounded by a lead shield box of 5 cm thickness, which is open from the top. The entire set up is kept inside an air-tight, temperature- controlled  cabin, to minimize the thermal electronic noise due to the ambient temperature variations.  Each minute, a data file is generated containing energy histogram (counts corresponding to different energies) accumulated over one -minute interval. The system is equipped to acquire an energy spectrum of photons ranging from $\sim$ 150 keV to $\sim$ 10 MeV. The energy spectra are calibrated regularly (weekly) using standard sources ($^{60}$Co and $^{137}$Cs) and background radioactivity peaks marked in Figure \ref{fig:one}.  More details of the  Equatorial secondary cosmic ray observatory (ESCRO) and experimental set-up can be found in \citet{vichare2018equatorial}. Atmospheric parameters, recorded using an in-house automatic weather station (AWS) and the India Meteorological Department (IMD-AWS) data at Tirunelveli, have been used in the present study. Vertical velocity (m/s) parameter is obtained by modifying vertical velocity (Pa/s) reanalysis data by Copernicus Atmosphere Monitoring Service Information [2017].
As mentioned in the previous section, SCR and the radioactivity of primordial radionuclides and their daughter nuclides have a significant contribution in the $\gamma$ flux measured on the ground. Therefore, in the energy spectrum obtained using NaI (Tl), a contribution from the $\gamma$ component of SCR is present along with distinct photopeaks due to background radioactivity over a continuum. For the convenience, the decay chains of $^{232}$Th, $^{235}$U, and $^{238}$U are displayed in the Appendix section. One photopeak of $^{40}$K, two of $^{208}$Tl, three of $^{214}$Bi, and one of $^{214}$Pb are clearly observed in the present data, as evident from Figure \ref{fig:one}. Among which, $^{40}$K is a primordial radionuclide, $^{208}$Tl is a daughter radionuclide of $^{232}$Th produced in the decay chain via $^{220}$Rn, while $^{214}$Bi and $^{214}$Pb are daughter radionuclides of $^{238}$U produced in the decay chain via $^{222}$Rn (Refer to the $^{232}$Th, $^{235}$U, and $^{238}$U decay chains visualised in the appendix). Except for $^{214}$Pb (energy 352 keV) that is too feeble in our spectrum, all the other observed photopeaks lie in between 500 keV -- 2.7 MeV.

\section{Observations}

\subsection{Diurnal pattern and its energy dependence}
\begin{figure*}
\vspace*{2mm}
\begin{center}
\includegraphics[width=14cm]{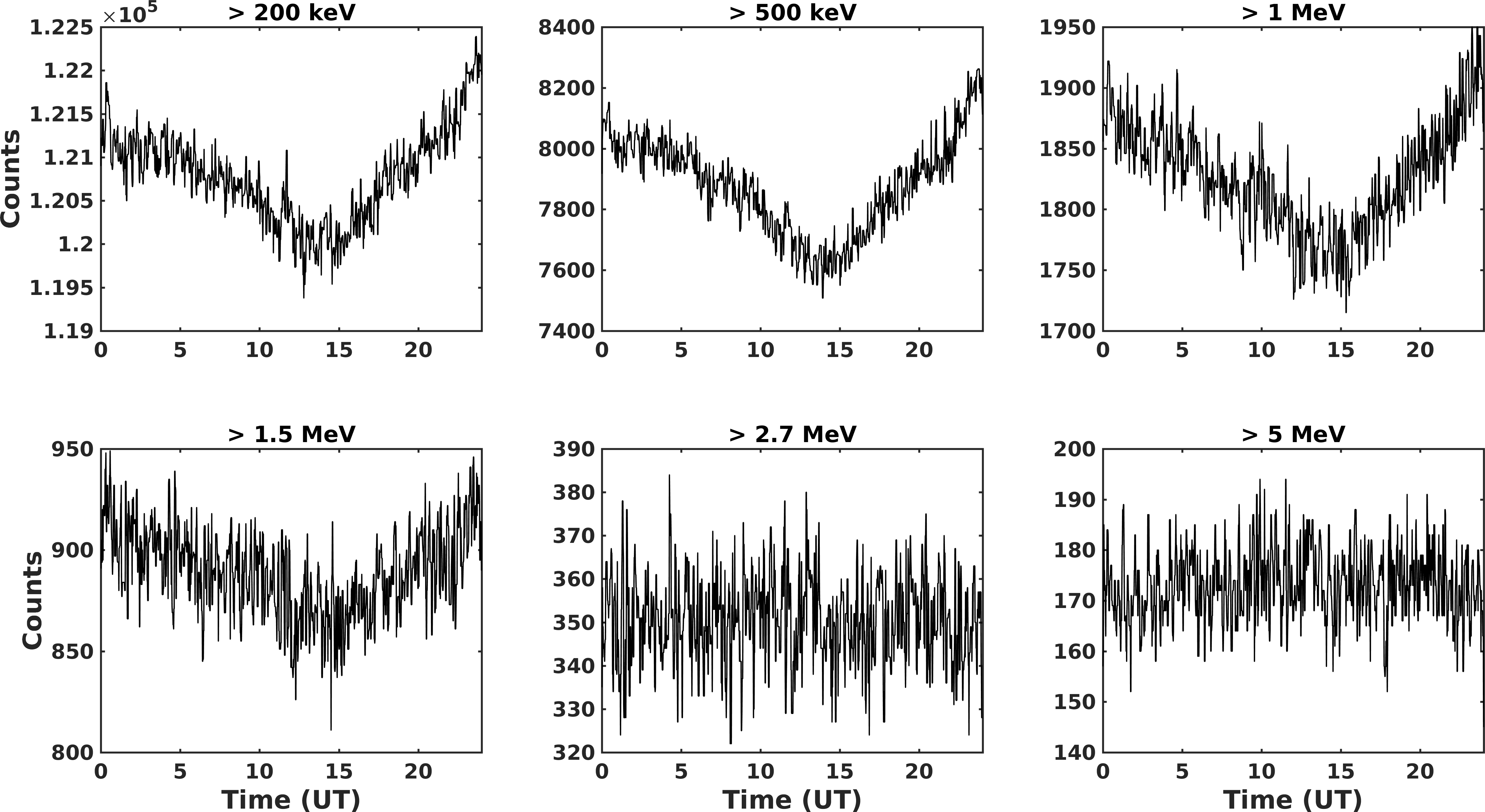} 
\end{center}
\caption{\textit{$\gamma$-ray data on 29 December 2017 is depicted for total six energy cut-offs ranging from $\sim$200 keV to $\sim$10 MeV plotted in subplots with energy as mentioned above each subplot. Number of counts accumulated in one minute are indicated on y-axes.}}
 \label{fig:two}
\end{figure*}

When total $\gamma$-ray counts accumulated over each minute is plotted against time during the day, it shows a distinct pattern similar to that reported by \citet{vichare2018equatorial}. The pattern shows a decrease in the $\gamma$-ray counts after the sunrise and recovery after the sunset. To study this variation with respect to energy, we applied filters with different cut-off energies. The application of a particular cut-off filter gives the total number of photons from all the energies above that particular energy, detected by the NaI (Tl) detector. Here, we have used 200 keV, 500 keV, 1 MeV, 1.5 MeV, 2.7 MeV, and 5 MeV as different energy cut-offs. The original spectrometer data of 1 min resolution is averaged over 5 min period. Temporal profiles of the counts with different energy cut-offs, on 29 December 2017 are shown in Figure \ref{fig:two}. 

As explained above, the counts with 200 keV cut-off contain particles with energies above 200 keV. Similarly, for subsequent panels, counts are shown for the corresponding cut-off energies. It is observed from the plots depicted in Figure \ref{fig:two} that a clear pattern is seen in the plots with cut-off energies from 200 keV to 1.5 MeV. Around 1-2 UT, counts start decreasing, and after 13 UT, the increase is started. Note that the local time corresponding to the onset of the decrease is near sunrise, while the beginning of the increasing trend corresponds to the local sunset.  The counts for energy cut-offs higher than 2.7 MeV do not show the pattern described above; instead, they show no variation with the time of a day, indicating the absence of any diurnal pattern for the energies above 2.7 MeV. 

\begin{figure*}
\vspace*{2mm}
\begin{center}
\includegraphics[width=8cm]{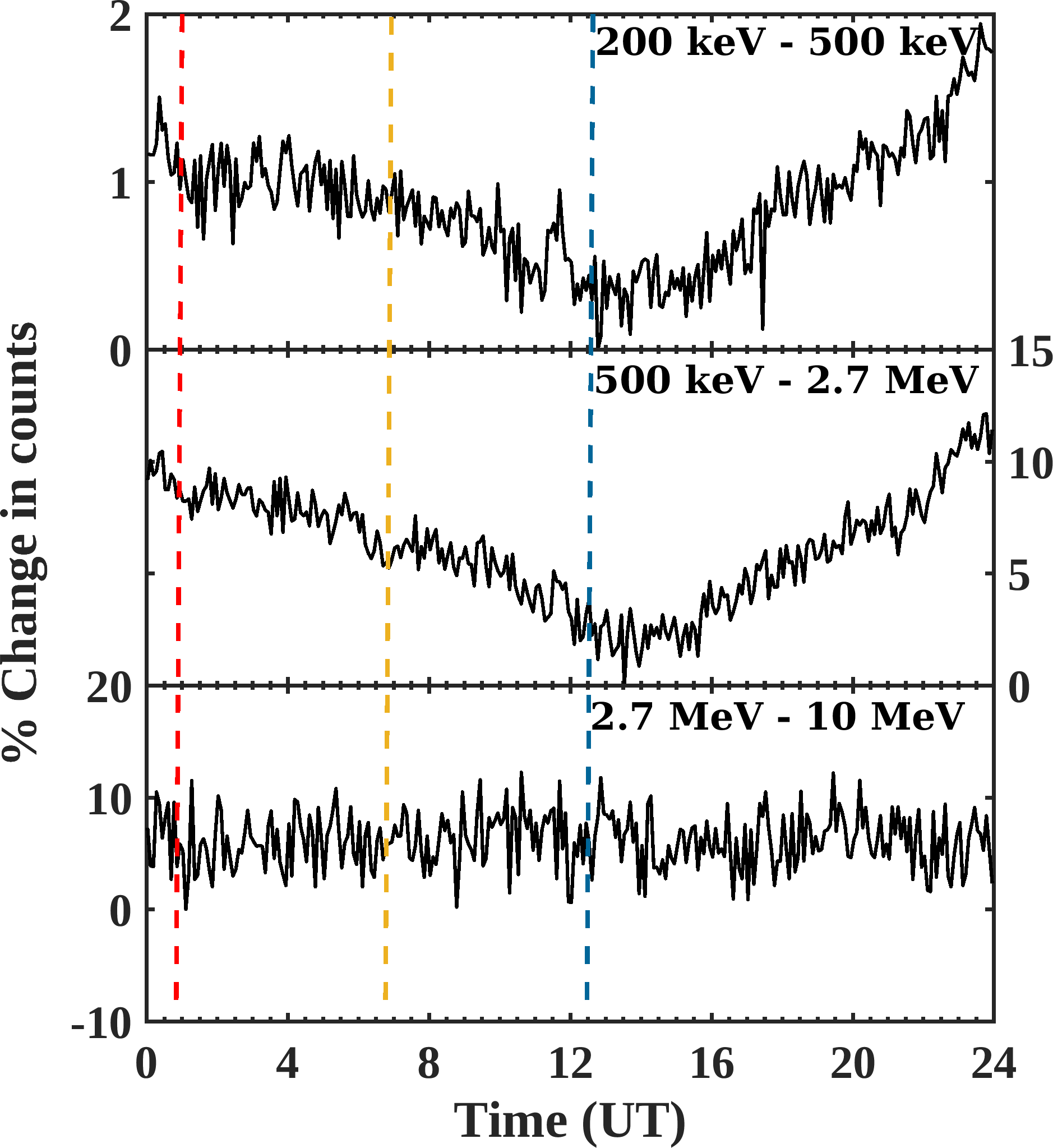} 
\end{center}
\caption{\textit{Temporal variation of $\gamma$-ray counts on 29 December 2017 for three different energy bands: 200 -- 500 keV, 500 keV -- 2.7 MeV, 2.7 -- 10 MeV. Variation in percentage in counts per minute are indicated on y-axes. Vertical red, blue, and yellow dashed lines indicate local sunrise, sunset, and noon times, respectively.}}
 \label{fig:three}
\end{figure*}

Therefore, it is appropriate to examine the variation in different energy bands or energy ranges. As mentioned above, the background radioactivity dominates in 500 keV -- 2.7 MeV, hence we have considered this range as one of the energy band. Energies lower than this band are grouped as a lower energy range, i.e., 200 keV -- 500 keV. The counts in the energy range between 2.7 -- $\sim$10 MeV is selected as a higher energy band. As the number of counts in an energy band varies due to the background radioactivity variations, it is appropriate to consider the percentage variations calculated with respect to the minimum count in that energy band. Figure \ref{fig:three} shows the diurnal variation of the \% change in counts in three energy bands. The amplitude of the diurnal variation is defined as the maximum \% variation during the day. 

In Figure \ref{fig:three}, the x-axis shows time in UT; the corresponding energy range is mentioned in each panel. We have shown sunrise and sunset (red and blue dashed lines) and solar noon (yellow dashed line) times at the observation site on the presented day (29 December 2017). It is observed that the pattern similar to that observed in Figure \ref{fig:two} that is present in the first two energy bands, 200 -- 500 keV and 500 keV -- 2.7 MeV. The \% of counts starts decreasing after around 1 UT and follows a decreasing trend up to around 13 UT. After sunset terminator, the counts start to increase, indicating recovery of the counts. It should be noted that though the pattern in the first two energy bands looks similar, there is a significant difference in the amplitude of the diurnal variation in both the energy bands. In the middle energy band (500 keV-2.7 MeV), it is about 12\%, while in the lower energy band, it is less than 2\%. This indicates that the major contribution to the diurnal pattern of $\gamma$-rays is coming from the middle energy band, which is primarily associated with the background radioactivity. There are a few $\gamma$-flux peaks, e.g., $^{214}$Pb (of energy 352 keV), along with the Compton component of other higher energy photopeaks due to background radioactivity in the lower energy band as well, which might have caused the observed diurnal pattern in that band. It can be observed from the bottom panel of Figure \ref{fig:three} that the higher energy band (2.7 -- 10 MeV) does not show any discernible diurnal pattern. Thus, it is observed that the significant diurnal variation exists only in the middle energy band, and hence, hereafter, the diurnal variations only for 500 keV -- 2.7 MeV energy band are presented.

\subsection{Seasonal variation}
In order to examine the robustness of the observed pattern on statistical grounds, three days from each month are selected for the year 2017, based on fair-weather conditions such that the maximum wind speed does not exceed 5 m s$^{-1}$.

\begin{table*}
\caption{List of fair weather days in 2017 used in the analysis}
\label{table:days}
\vskip4mm
\centering
\begin{tabular}{lccc}
\hline
Month & Day 1 & Day 2 & Day 3\\
\hline
January & 10 & 14 & 23\\
February & 05 & 11 & 18\\
March & 13 & 24 & 28\\
April & 01 & 08 & 27\\
May & 01 & 05 & 06\\
June & 08 & 10 & 22\\
July & 22 & 25 & 29\\
August & 03 & 21 & 27\\
September & 03 & 21 & 30\\
October & 02 & 17 & 20\\
November & 11 & 21 & 24\\
December & 11 & 28 & 29\\
\hline
\end{tabular}
\end{table*}

\begin{figure*}
\begin{center}
\includegraphics[width=17cm]{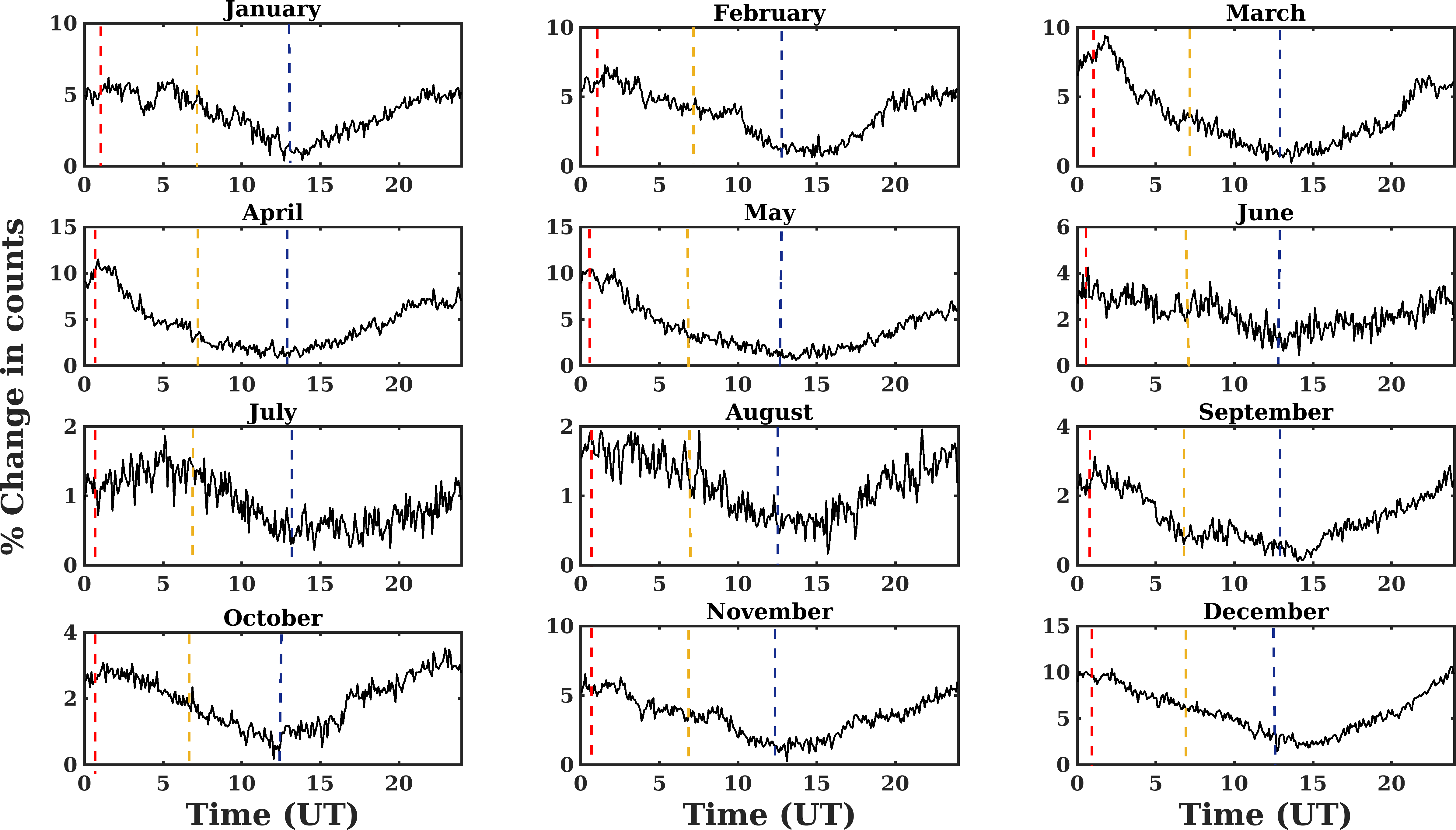} 
\end{center}
\caption{\textit{Each panel depicts the average of \% variation of $\gamma$-ray counts with energies between 500 keV -- 2.7 MeV on three different days in each month from January to December 2017. Vertical red, blue, and yellow dashed lines indicate local sunrise, sunset, and noon times, respectively.}}
 \label{fig:four}
\end{figure*}

\begin{figure*}
\vspace*{2mm}
\begin{center}
\includegraphics[width=8cm]{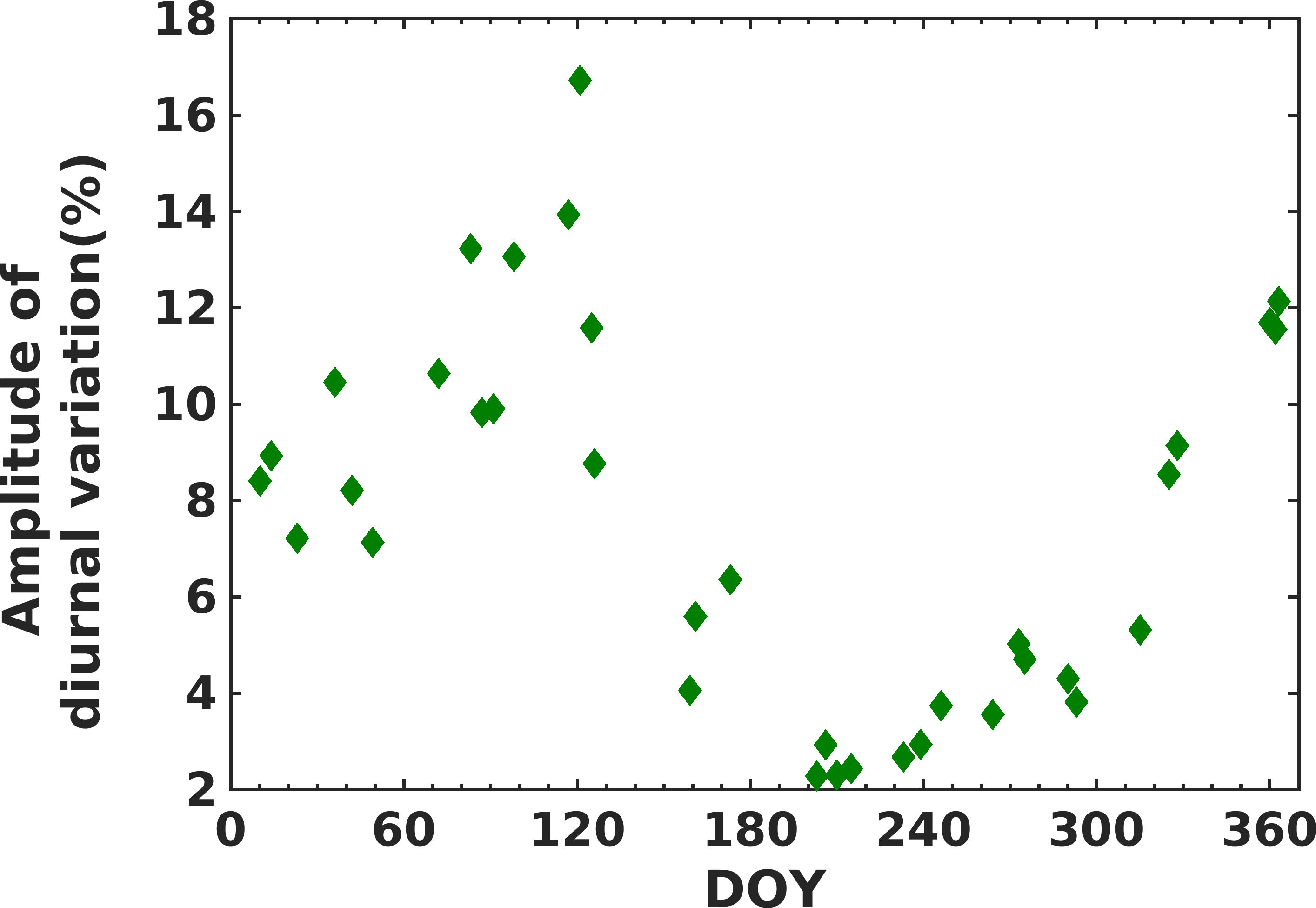} 
\end{center}
\caption{\textit{Observed diurnal variation of the $\gamma$-ray flux on individual days; x-axis shows day of year (DOY)}}
 \label{fig:five}
\end{figure*}

Figure \ref{fig:four} shows the daily variation in the energy band 500 keV -- 2.7 MeV for the months January to December 2017. Each subfigure depicts the averaged diurnal pattern on three different days, indicated in Table \ref{table:days}. It is observed that in January and February, the $\gamma$-ray counts on individual days follow the pattern of decrease around 1 UT and increase after around 13-14 UT. The amplitude of variation in January and February is seen to be around 5.6\% and 7.2\%, respectively. For March, April, and May 2017, a range of variation is 9.2\%, 10.3\%, 10\%, respectively. In June 2017, the variation fell to 3.7\%, while in July and August, the pattern becomes very weak with just 1.8\% and 1.7\% variation, respectively. In September and October, it slightly increases to 3.1\% and 3.2\%, respectively. Although the range of variation is small for months from June to October, the diurnal pattern is still similar. In November, the variation increases to 6.4\%; for December, it further increases to 9.5\%. To visualise the trend of change in percentage variation, Figure \ref{fig:five} shows the range/amplitude of the diurnal variation on individual days (Listed in Table 1) against the day of the year (DOY). It is clear from the figure that the amplitude of diurnal variation is minimum in June-October months and maximum in April-May months. Thus $\gamma$-flux recorded on the ground exhibits a considerable seasonal dependence.

\section{Analysis of individual background photopeak} %to identify the cause
\begin{figure*}
%\vspace*{2mm}
\begin{center}$
\begin{array}{cc}
\includegraphics[width=5.3cm]{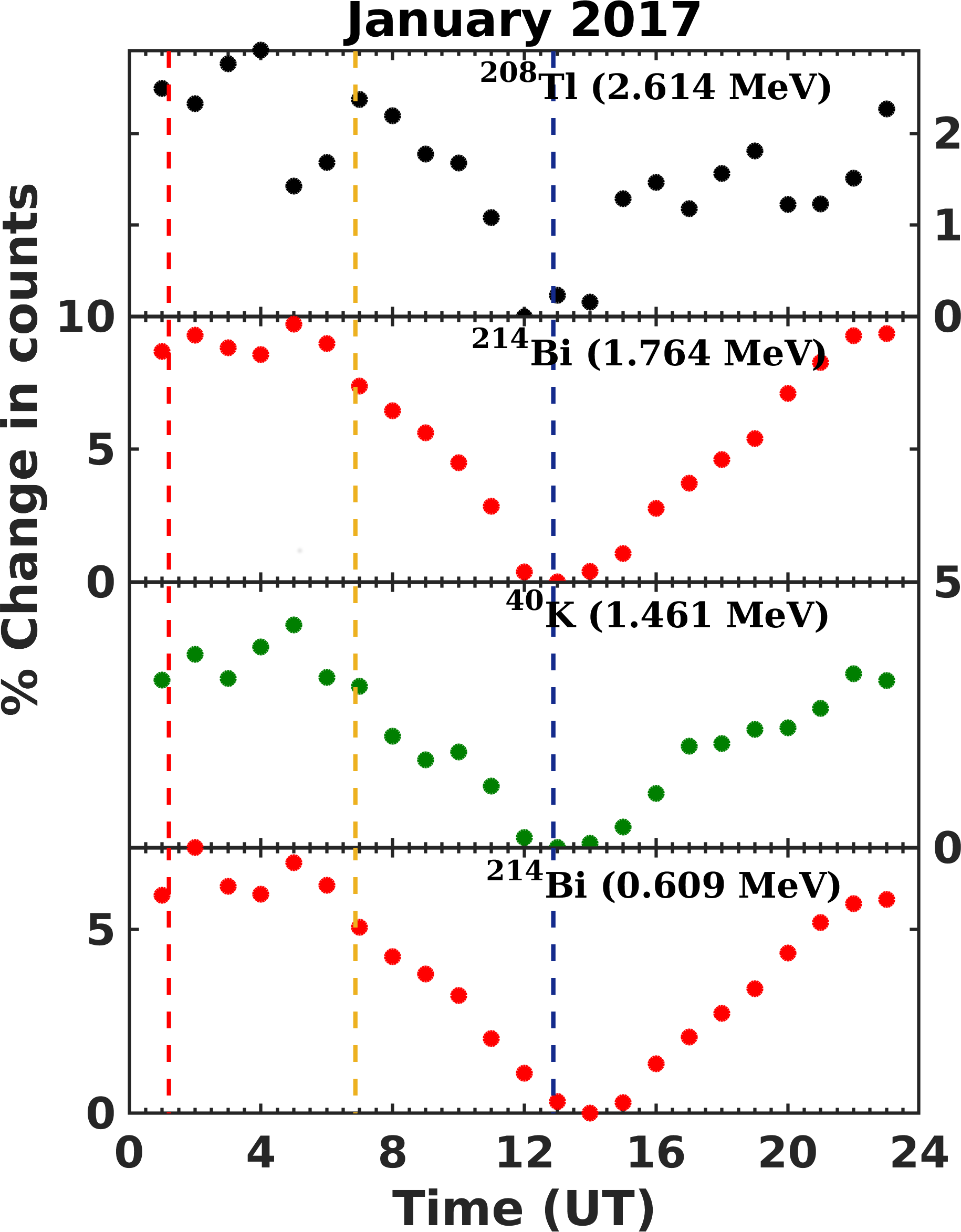} &
\includegraphics[width=5cm]{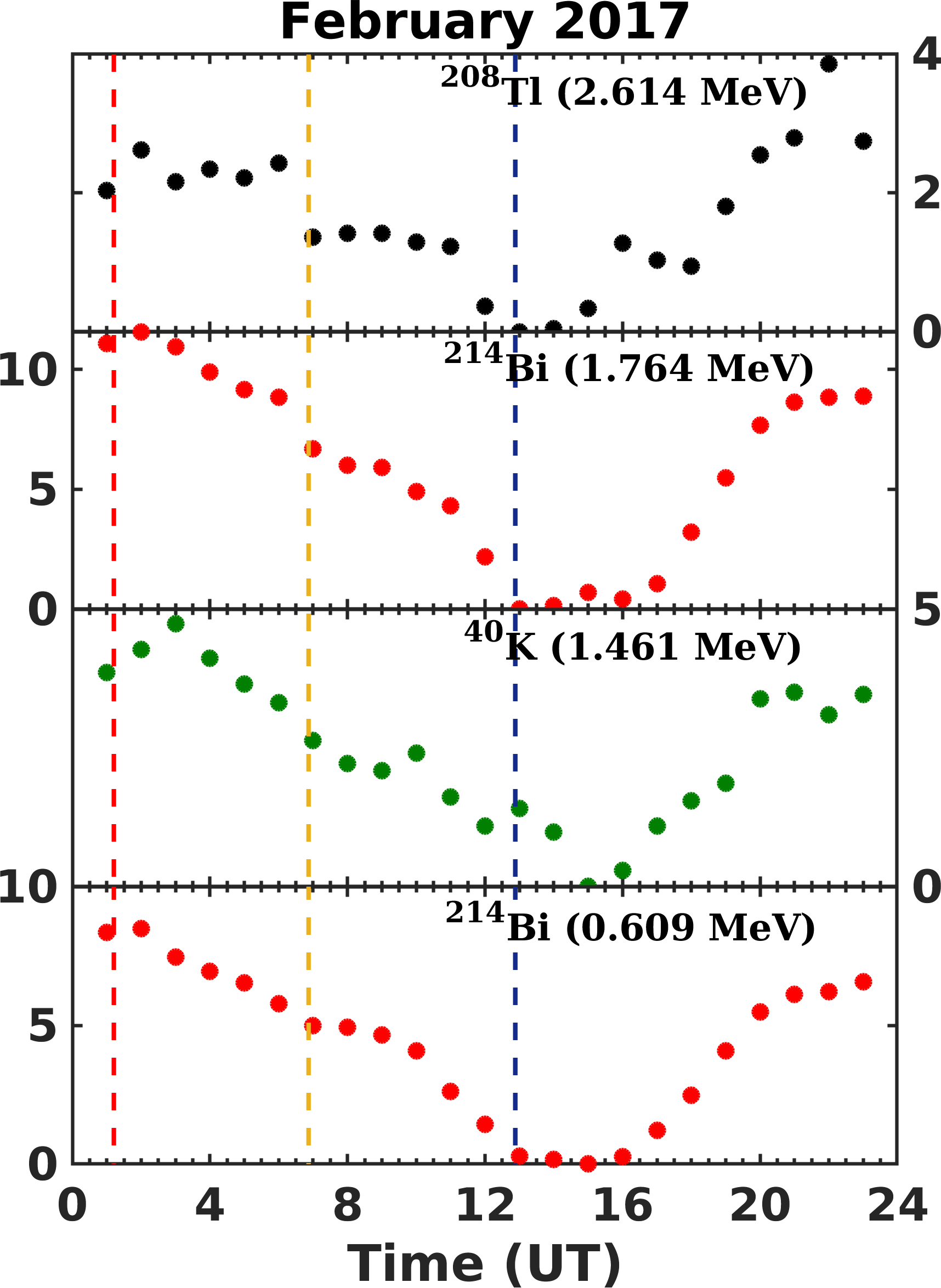}\\
\\
\includegraphics[width=5.45cm]{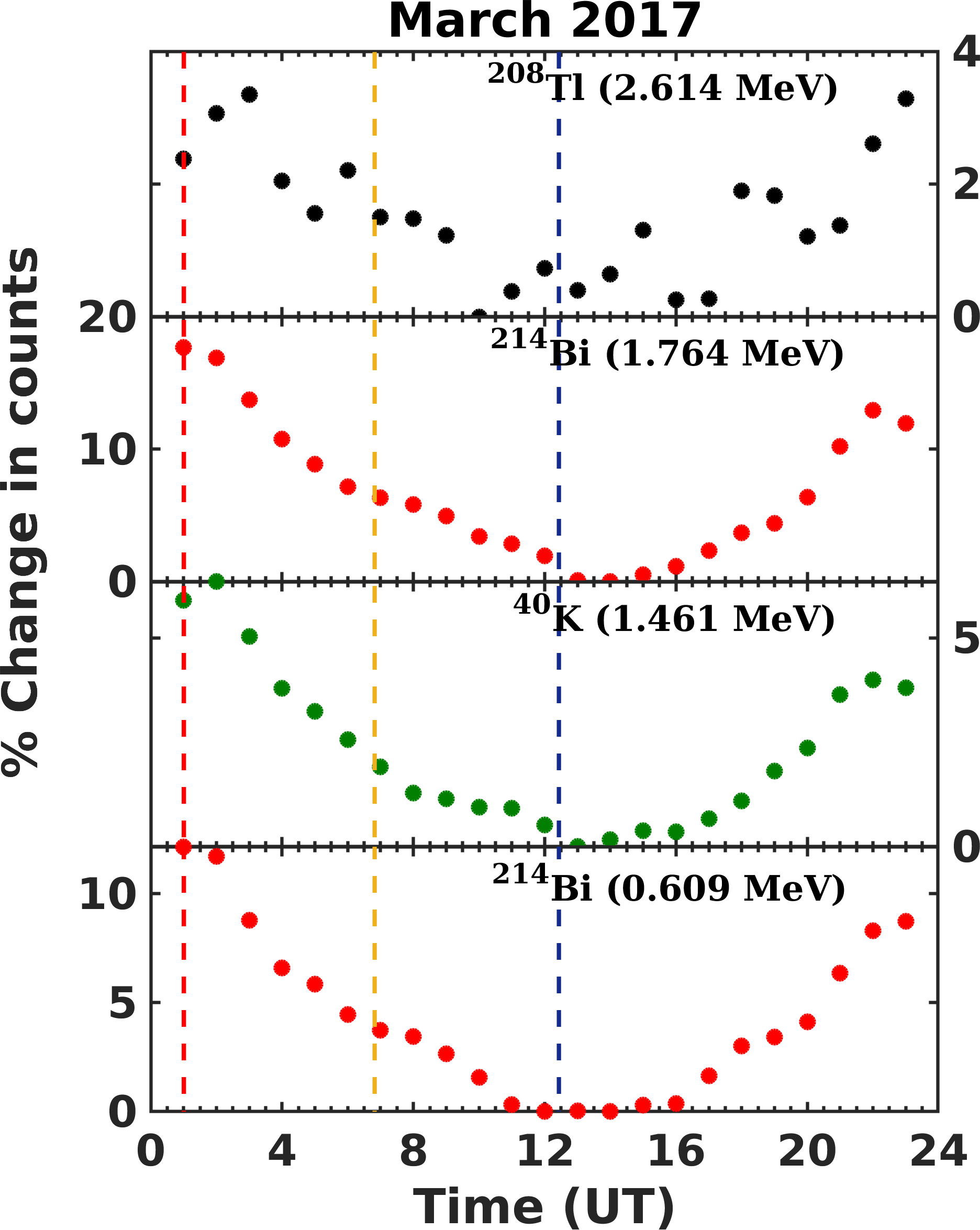} &
\includegraphics[width=5cm]{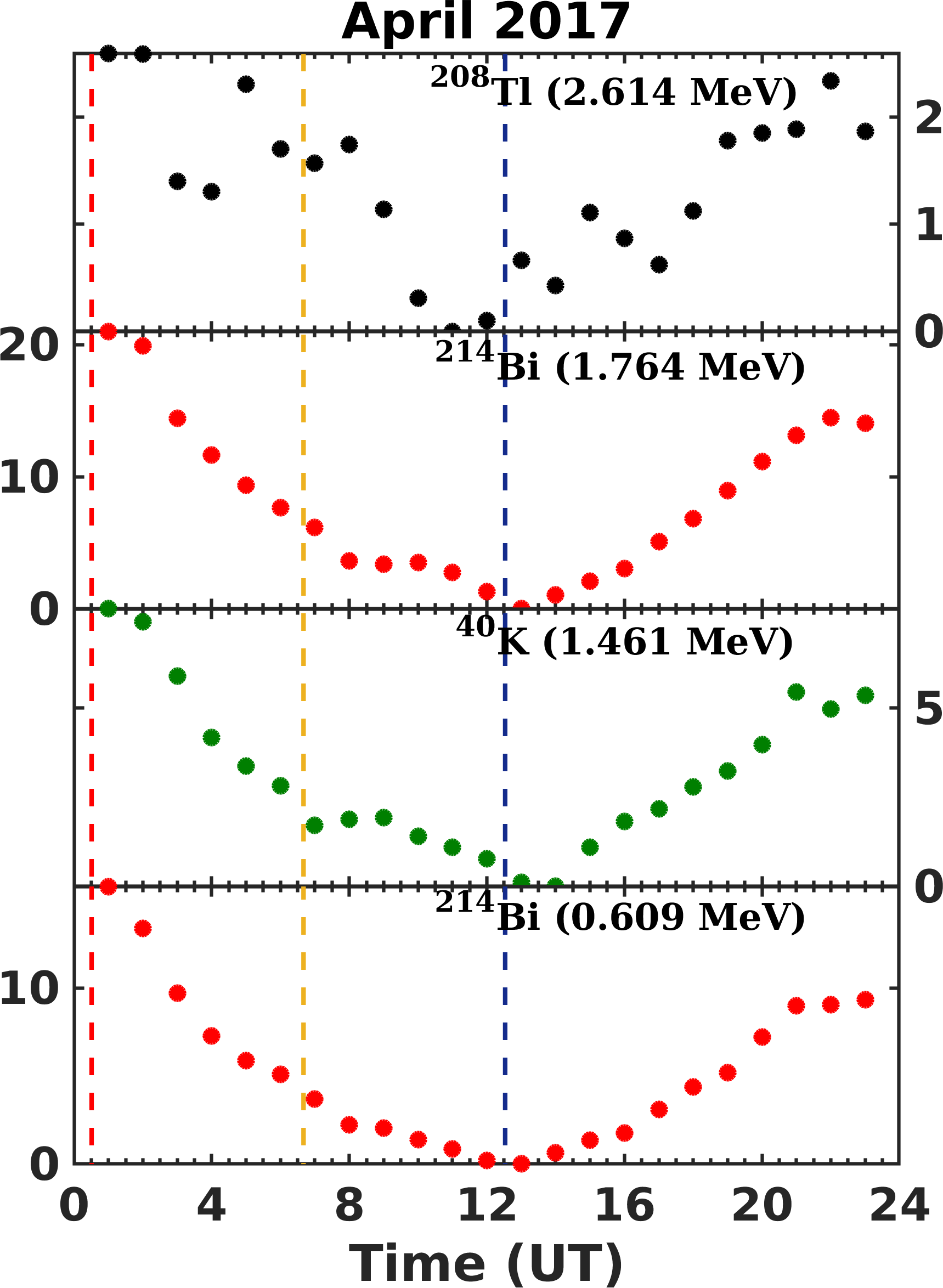}\\
\end{array}$
\end{center}
\caption{\textit{The diurnal pattern1 (average of three days) of (i)   $^{208}$Tl (2.614 MeV), (ii) $^{214}$Bi(1.764 MeV), (iii) $^{40}$K (1.461 MeV), and (iv) $^{214}$Bi(0.609 MeV) for each month from January to April 2017.}}
 \label{fig:six}
\end{figure*}

\begin{figure*}%\vspace*{2mm}
\begin{center}$
\begin{array}{cc}
\includegraphics[width=5.47cm]{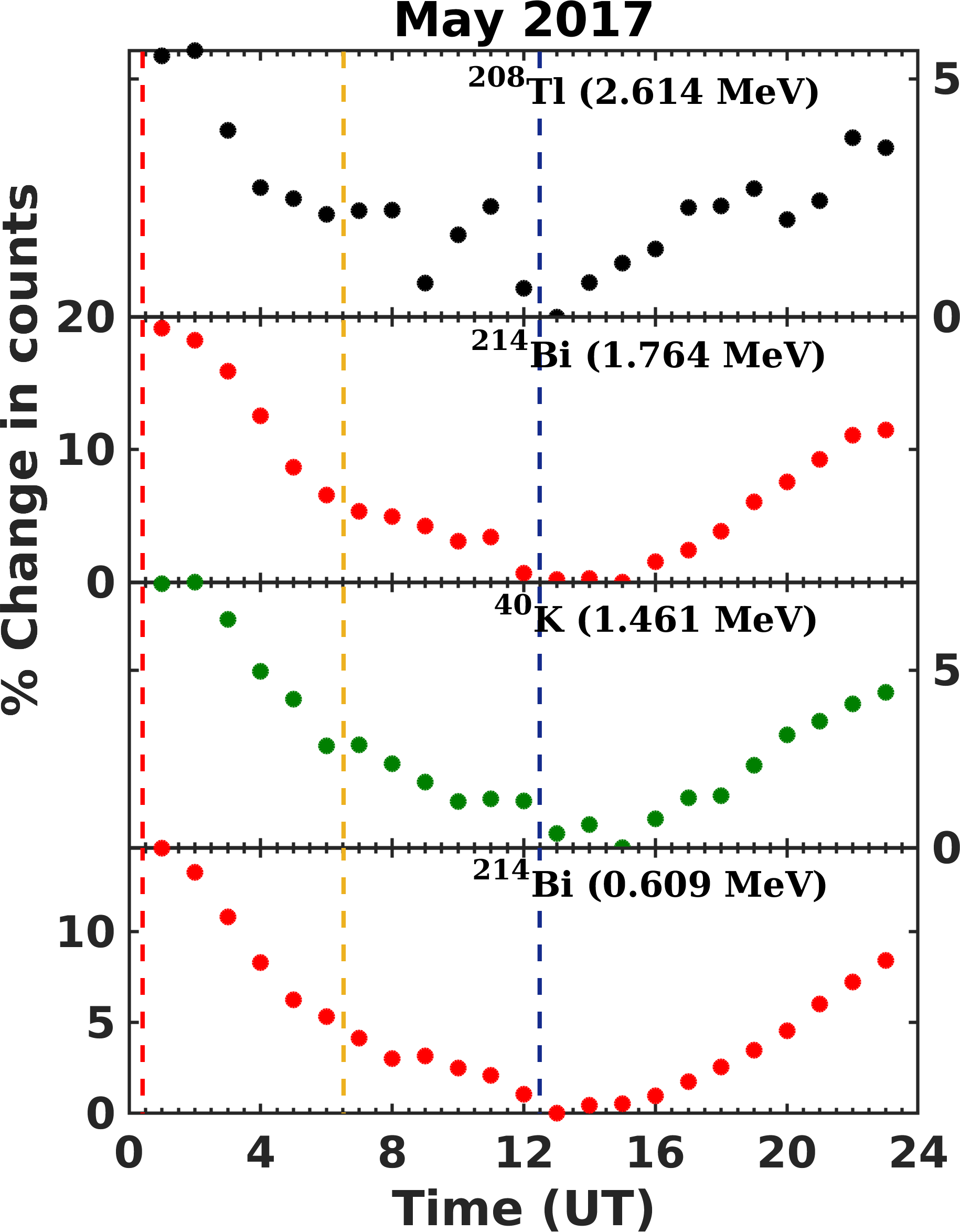} &
\includegraphics[width=5cm]{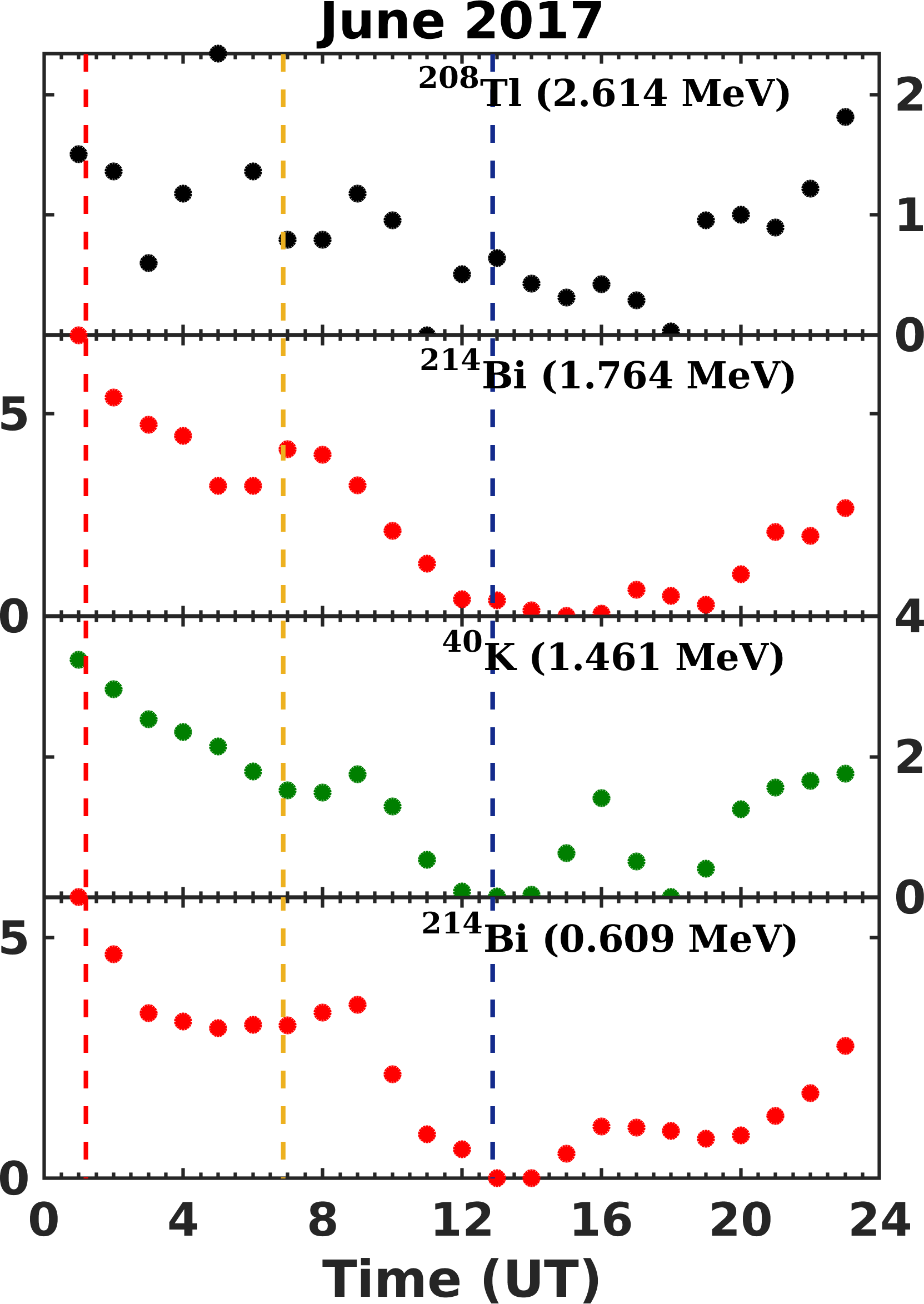}\\
\\
\includegraphics[width=5.42cm]{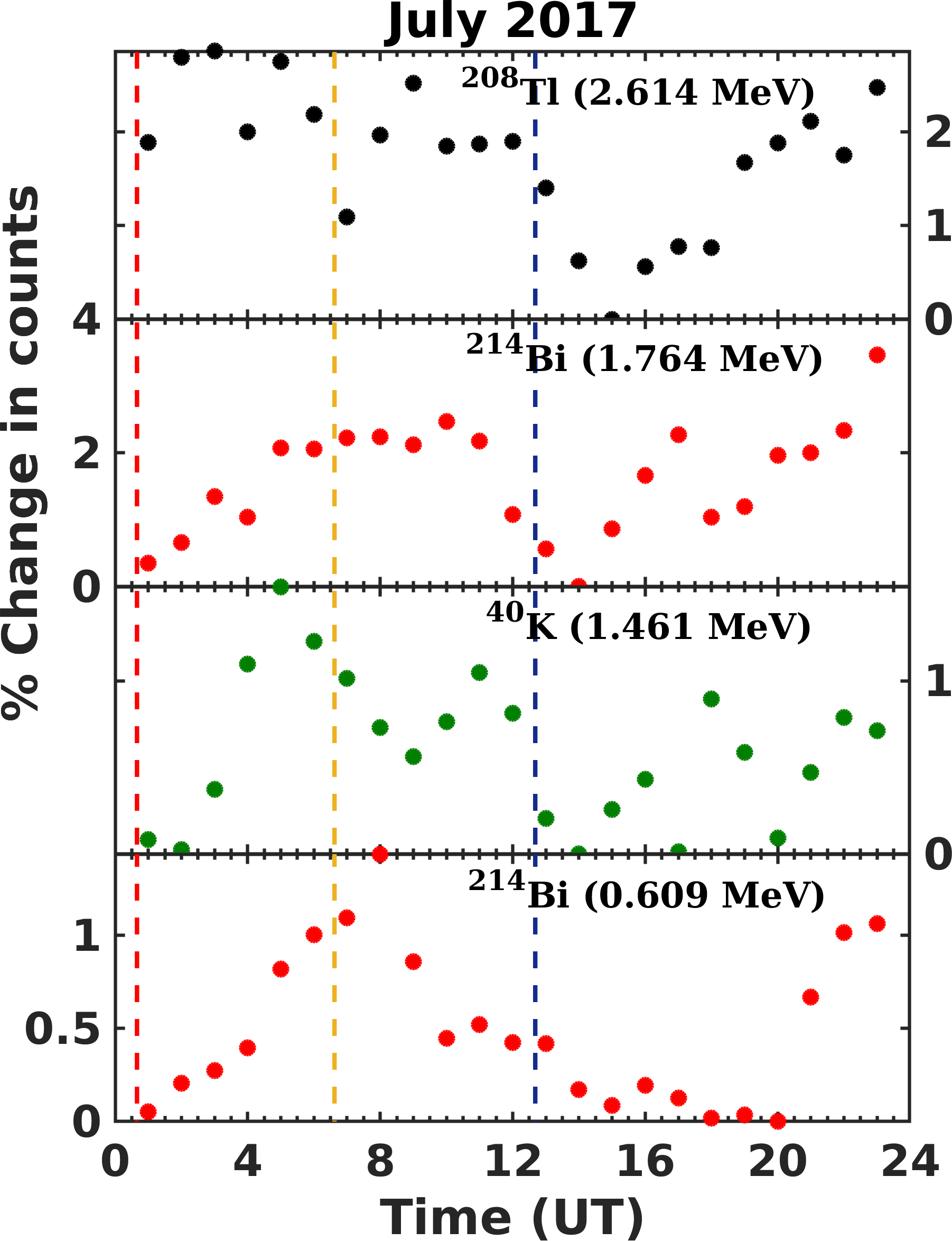} &
\includegraphics[width=5cm]{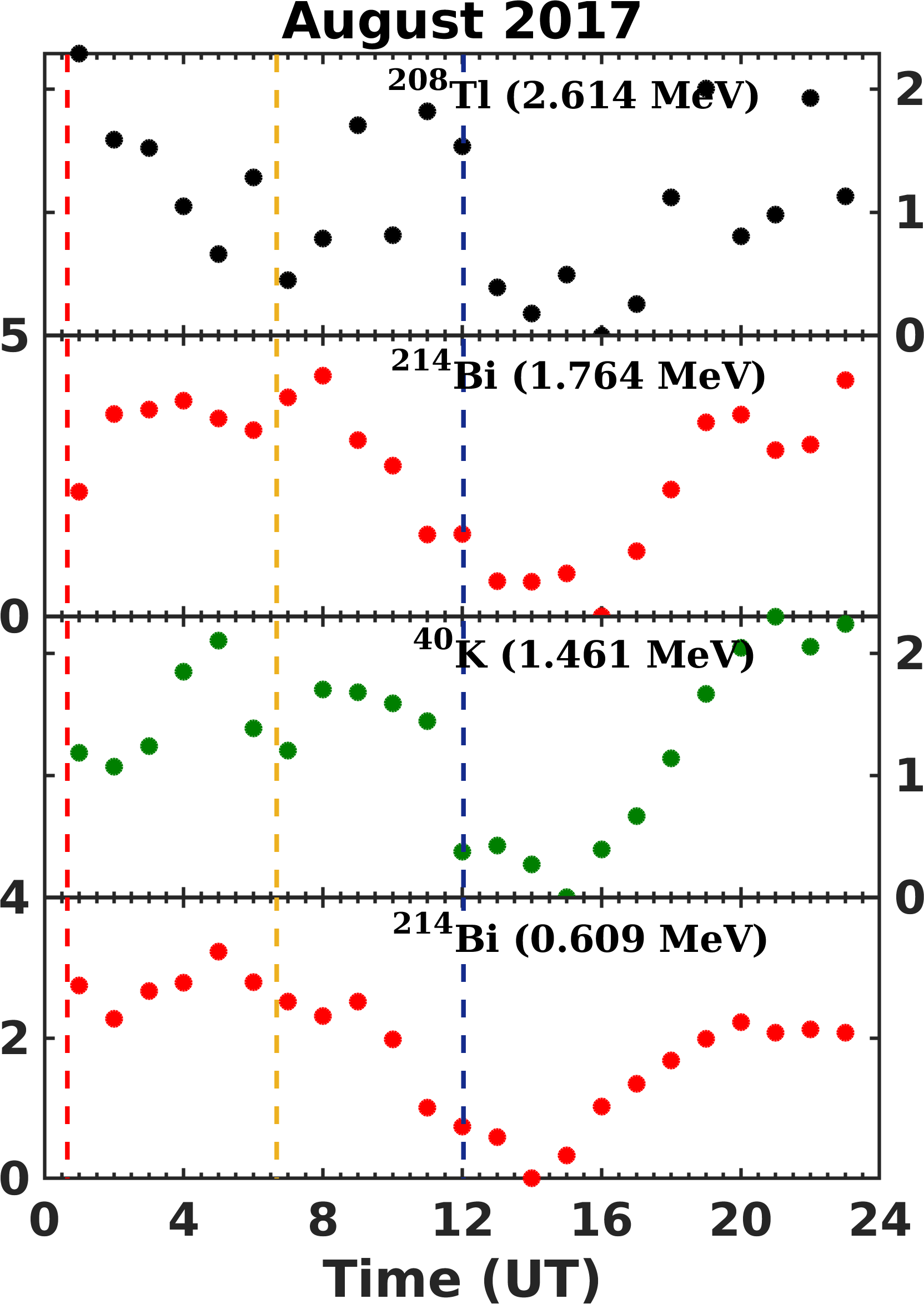}\\
\end{array}$
\end{center}
\caption{\textit{Same as Figure 6, but for each month from May to August 2017.}}
 \label{fig:seven}
\end{figure*}

\begin{figure*}
%\vspace*{2mm}
\begin{center}$
\begin{array}{cc}
\includegraphics[width=5.45cm]{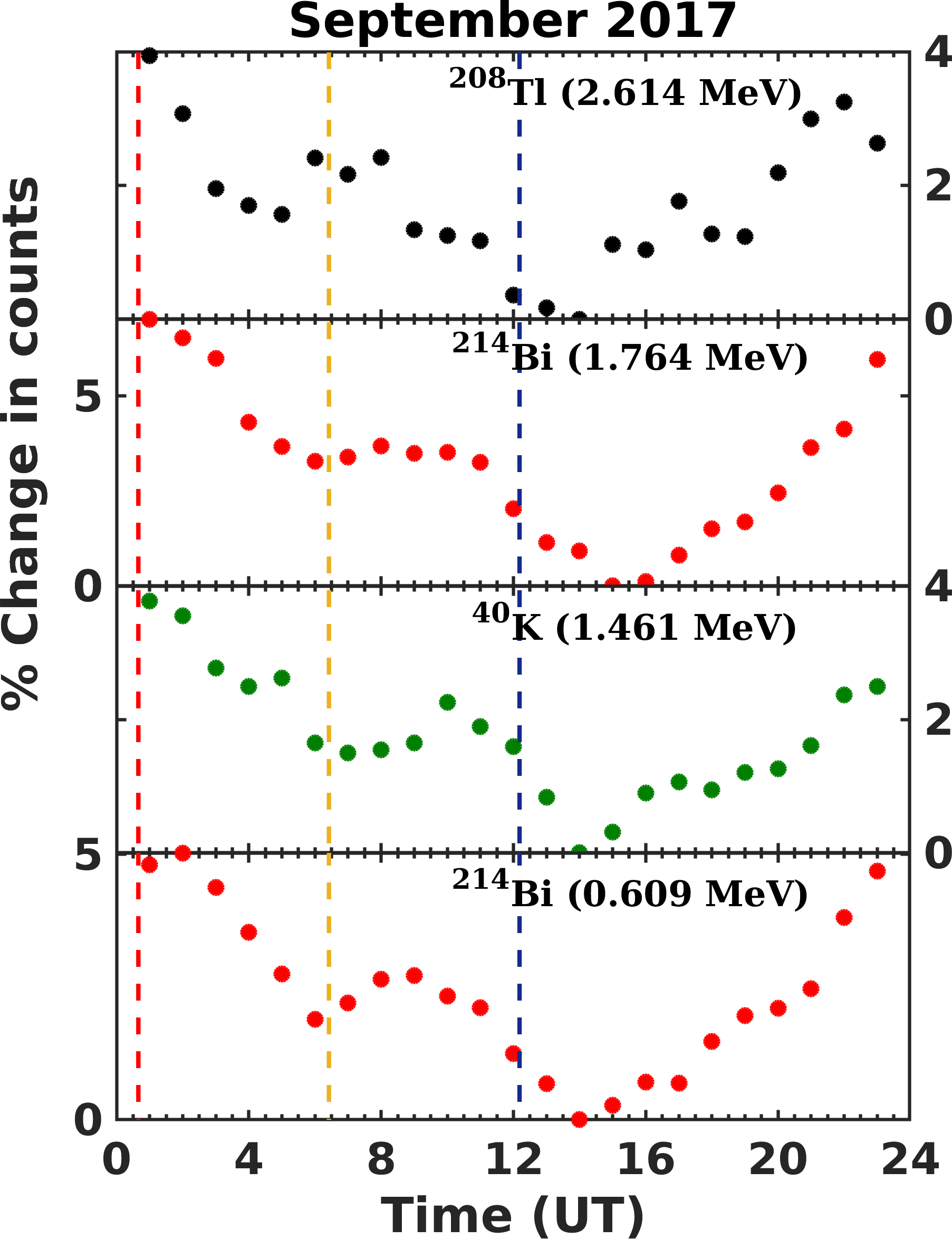} &
\includegraphics[width=5.05cm]{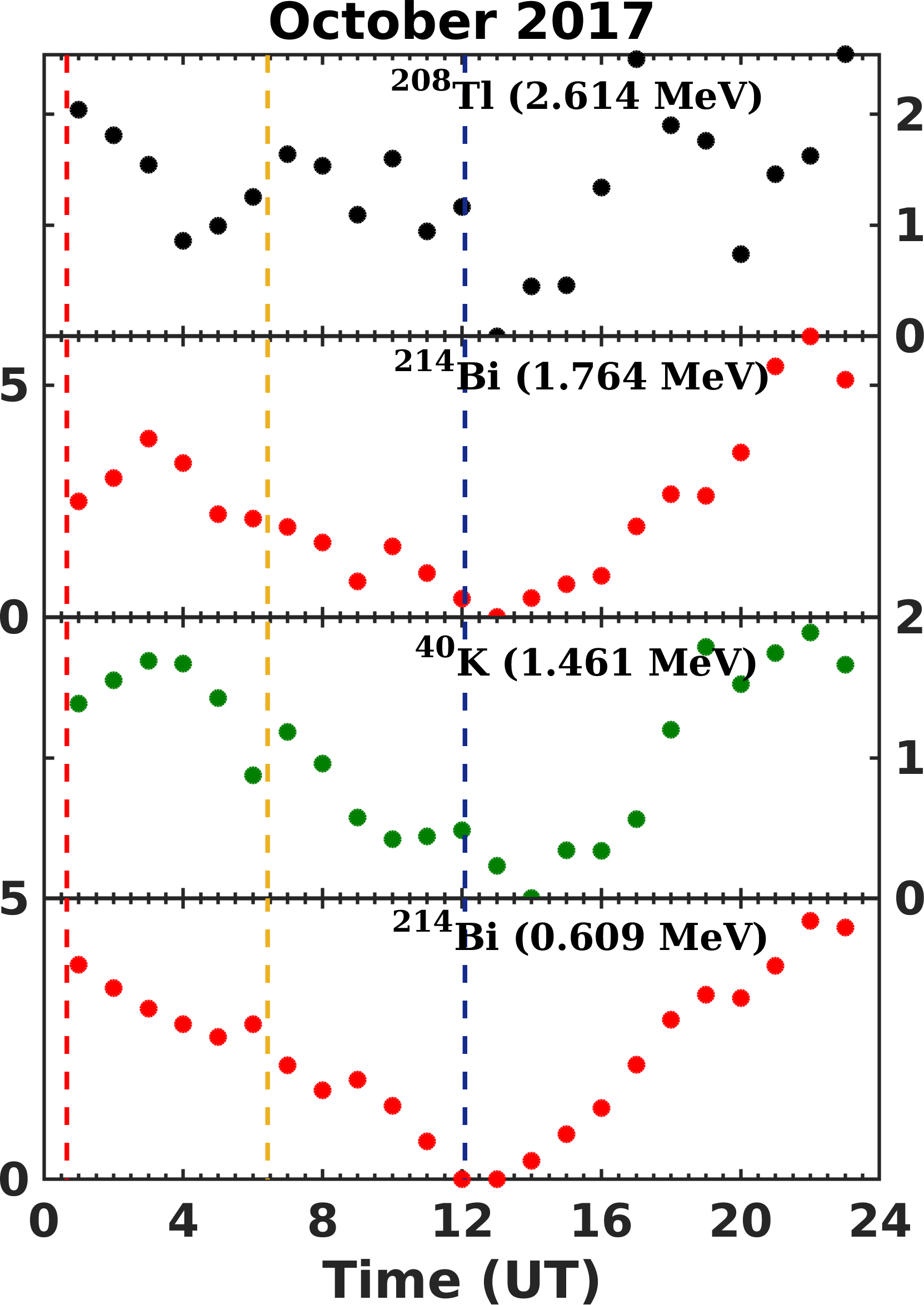}\\
\\
\includegraphics[width=5.45cm]{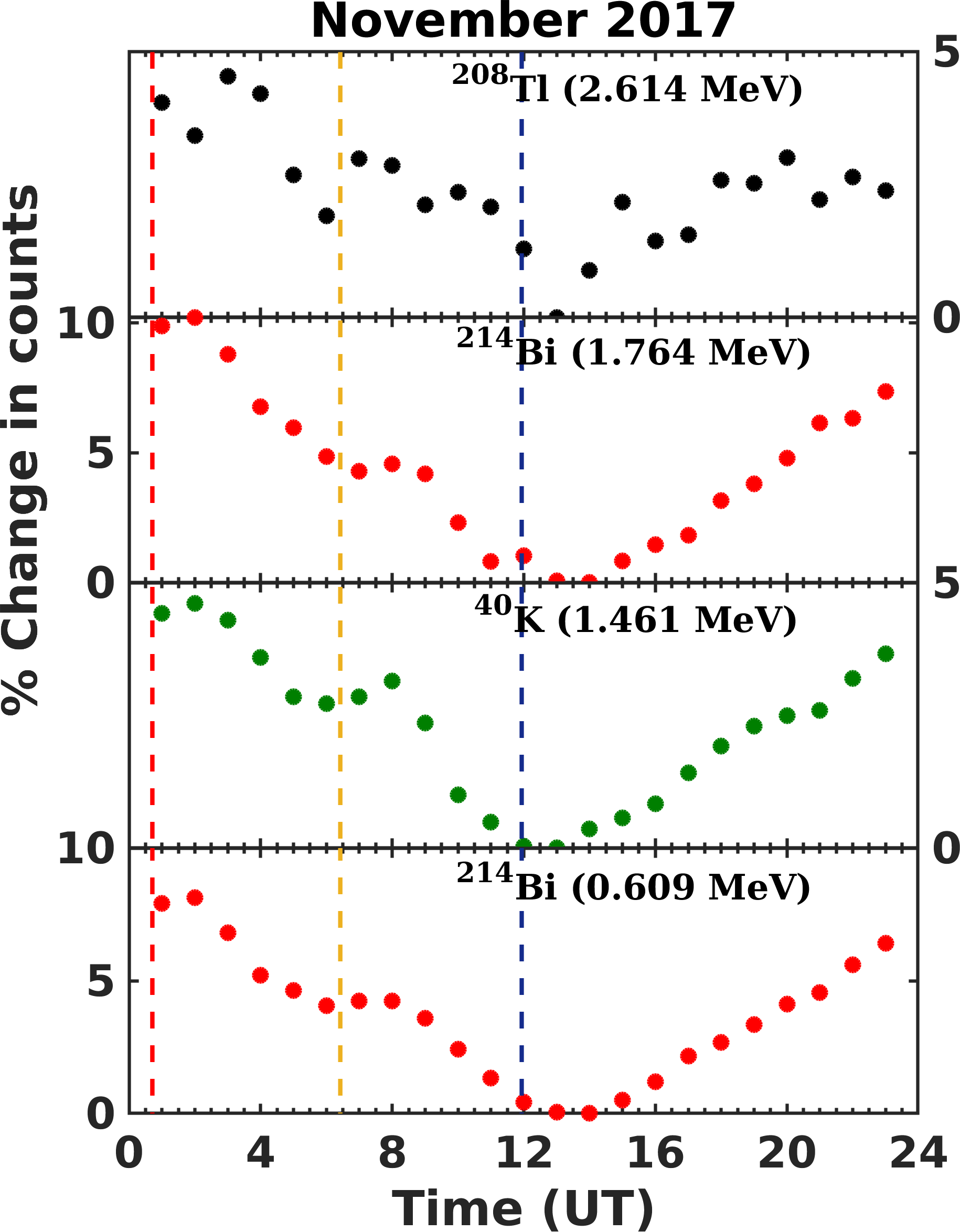} &
\includegraphics[width=5.15cm]{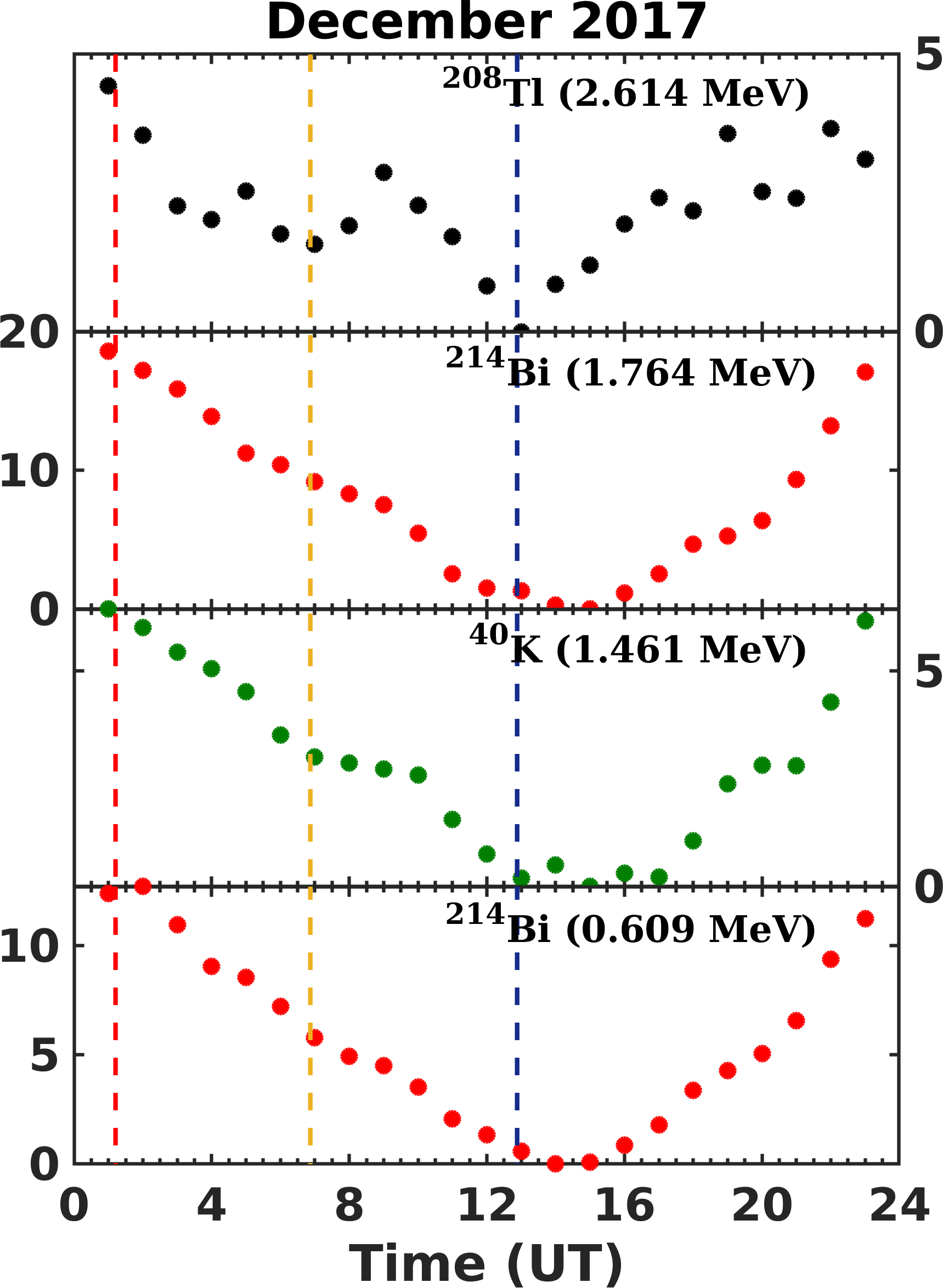}\\
\end{array}$
\end{center}
\caption{\textit{Same as Figure 6, but for each month from September to December 2017.}}
 \label{fig:eight}
\end{figure*}

It is evident from Figure \ref{fig:three} that the diurnal variation is most prominently seen in 500 keV -- 2.7 MeV energy band. As background radioactivity is dominant in this energy range, one can examine whether any particular radioactive elements are causing the observed pattern. In order to inspect this, we have estimated the counts in different photopeaks of radioactivity by calculating the area under the photopeak and checked whether it exhibits any diurnal pattern.

As discussed in \lq{Data description}' (Section 2), there are several photopeaks between 500 keV -- 2.7 MeV. To identify the peaks clearly, we have accumulated data for two hours. For the selected days, total flux is calculated (accumulated counts) with a two-hour window at each hour. Thus, the two-hour sliding window has a one-hour overlap so that 23 data points per day are obtained. Following  four photopeaks are chosen for further analysis, as they are significant and very clear in the spectrum: $^{214}$Bi (0.609 MeV), $^{40}$K (1.461 MeV), $^{214}$Bi(1.764 MeV), $^{208}$Tl (2.614 MeV). By calculating the area under a photopeak using the trapezoidal method of integration, we estimated the number of photons emitted by the element corresponding to that particular photopeak.

\begin{figure*}
\vspace*{2mm}
\begin{center}
\includegraphics[width=10cm]{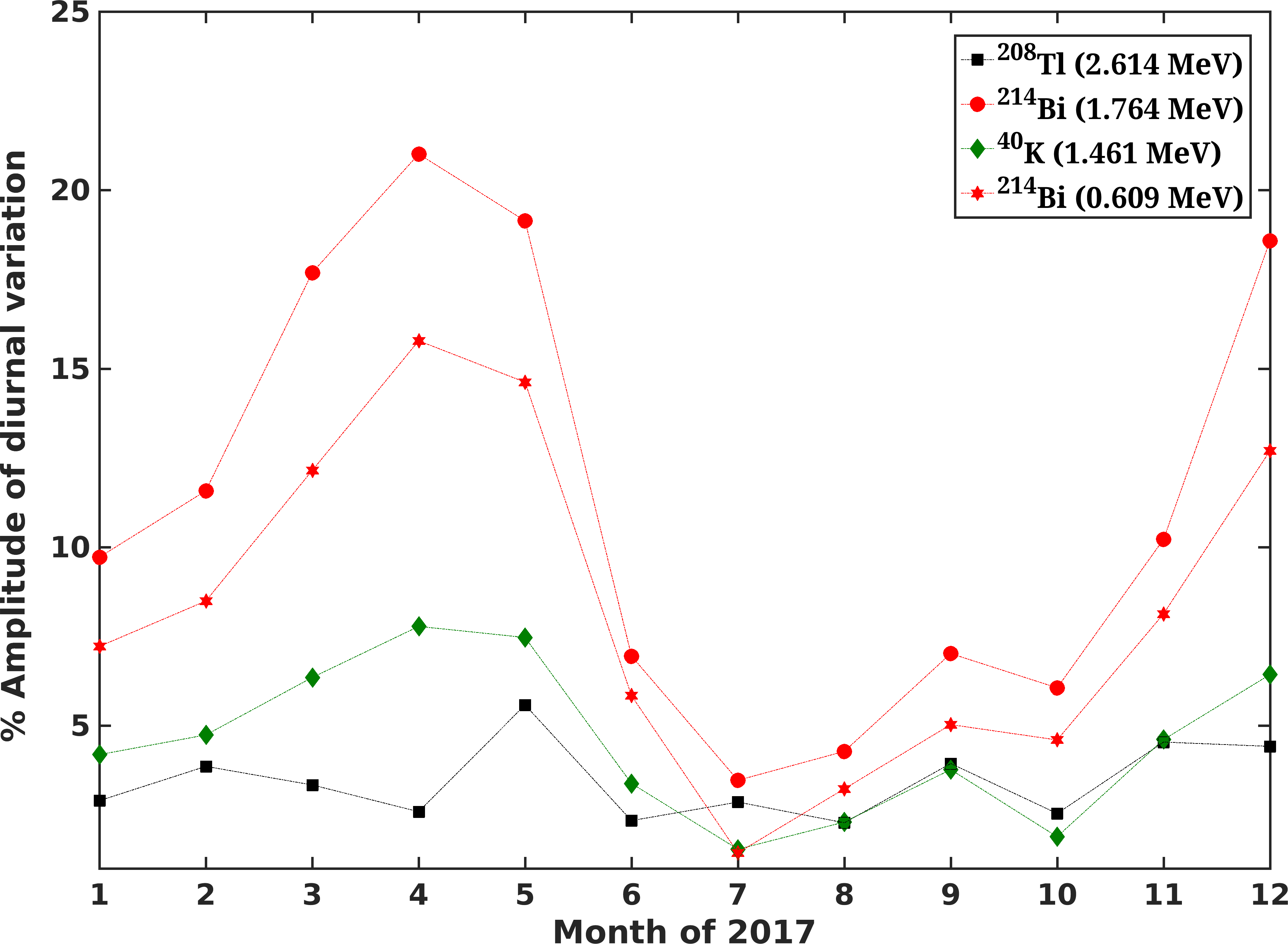}
\end{center}
\caption{\textit{Seasonal variation in the amplitude of individual photopeaks: x-axis indicates month of year 2017; y-axis indicates the amplitude of average diurnal variation in counts under the photopeak in \%.}}
 \label{fig:nine}
\end{figure*}

Figures \ref{fig:six}, \ref{fig:seven}, and \ref{fig:eight} together depict the variation in the counts under the photopeaks of radionuclides $^{214}$Bi (0.609 MeV, 1.764 MeV), $^{40}$K (1.461 MeV), and $^{208}$Tl (2.614 MeV). The mean of three days' counts for a month is presented in a subfigure for the annotated radionuclides. It is observed from the plots that the temporal variation of the counts under almost all the selected photopeaks shows a pattern similar to that described in the previous section. The pattern is quite robust and intense for two peaks of Bi and one peak of K, but not very clear in some cases for $^{208}$Tl. But more or less, a similar diurnal pattern is observed in all the selected photopeaks. It is noted that the percent change in the counts varies for different photopeaks in a given month. The amplitude for two $^{214}$Bi  peaks is quite large (e.g., $\sim$ 15-20\% during March-April months) and that for $^{208}$Tl is very small ($\sim$ 3-4\%) as compared to other peaks. The diurnal amplitude for $^{40}$K is intermediate ($\sim$ 6-8\%).
From Figures \ref{fig:six}, \ref{fig:seven}, and \ref{fig:eight}, it can be observed that the percent change in the counts of each photopeak varies from month to month in year 2017. Except for $^{208}$Tl, the other three photopeaks show maximum variation during April, and the amplitudes are lowest during June-October.  In June, $^{214}$Bi(1.764 MeV) shows 7\%, $^{40}$K shows 3\%, and $^{214}$Bi (0.609 MeV) shows 6\% variation. The diurnal pattern also seems to deviate from the general trend in the months of June as well as July.  In July, $^{214}$Bi(1.764 MeV) shows 3.5\%, $^{40}$K shows 1.5\%, and $^{214}$Bi (0.609 MeV) shows 1.5\% variation. The diurnal pattern is significantly disturbed in the month of July for all the photopeaks. The pattern is restored from October onwards. For $^{208}$Tl peak, the daily range of variation is $\sim$ 3-4\% and do not show any seasonal dependence. The seasonal variation in each photopeak is depicted in Figure \ref{fig:nine}. It is evident from the figure that the $^{214}$Bi (1.764 MeV) has the strongest seasonal dependence, $^{40}$K has moderate dependence, and  $^{208}$Tl has almost no seasonal dependence. Notably, an increase in the month of April is a feature present in all four photopeaks. It can be remarked that the seasonal dependence of the $^{214}$Bi peaks is similar to that of $\gamma$-flux depicted in Figure \ref{fig:five}, indicating that the diurnal pattern observed in the $\gamma$-flux is mainly due to the $^{214}$Bi radionuclide. 

%\pagebreak
\section{Discussion}
The present work investigates the diurnal variation of the $\gamma$-ray spectrum recorded by the NaI (Tl) scintillation detector. It is demonstrated here that there exists a diurnal pattern in the $\gamma$-ray flux measured near the ground. The $\gamma$-ray counts start decreasing at a time coinciding with local sunrise, and the decrease is observed up to local sunset, after which, the counts start increasing. This diurnal pattern is observed almost on all fair weather days throughout the year, thus indicating robustness. There are few earlier works reporting the similar diurnal trend noticed in the $\gamma$-rays detected by Ge(Li) detector \citep{katase1982variation} and NaI(Tl) detector \citep{raghav2013confirmation, raghav2015low, vichare2018equatorial}. The present study carried out a detailed analysis of this pattern and attempted to identify the possible responsible factors. 

In this context, we examined the pattern of the $\gamma$-flux of different energies. As mentioned in the introduction, the $\gamma$-rays observed on the ground can have contribution due to background radioactivity and secondary particles produced in the cosmic ray interaction with the atmospheric particles. The terrestrial background radioactivity is, in general, observed in the energy range between 500 keV to 2.7 MeV. Therefore, it is appropriate to inspect the response in different energy ranges to identify the responsible contributor in the observed diurnal pattern. It is observed that a strong diurnal pattern is unambiguously present in the 500 keV-- 2.7 MeV energy range, and completely absent in the higher energies ($>$ 2.7 MeV). As the $\gamma$-rays (photons) due to cosmic ray interactions can exist in the entire 100 keV-- 10 MeV energy range, it can be concluded that the observed diurnal variation in the $\gamma$-ray flux could be only due to the background radioactivity. 
Normally, in the energy spectrum collected at the observation site, four very distinct peaks, including $^{40}$K (1.461 MeV), $^{208}$Tl (2.614 MeV), and two peaks of $^{214}$Bi (0.609 MeV, 1.764 MeV) are clearly visible. We found that the diurnal variations of the counts present in these photopeaks exhibit a diurnal pattern similar to that observed for the $\gamma$-flux. The pattern is quite robust and intense for the two peaks of $^{214}$Bi and $^{40}$K, whereas a weak pattern is observed for $^{208}$Tl. Roughly a similar diurnal pattern is observed in all the selected photopeaks.

Then the next question is what is the source for the observed diurnal pattern in the $\gamma$-ray flux. \citet{katase1982variation} ascribed the diurnal variation of $\gamma$-ray intensity to the fluctuations in the atmospheric concentration of $^{222}$Rn, which induce the variation of concentrations of $^{214}$Pb and $^{214}$Bi components in the atmosphere, resulting in the variation of $\gamma$-ray flux. They attributed diurnal variation to the generation of \lq{inversion layer}' and its dissipation after sunrise. Their findings were based on the data collected on some days during three months (January, July, and November). There are many papers reporting the diurnal variation in the concentration of \textit{radon} [\citet{el2001diurnal, schubert2002diurnal, desideri2006monitoring, galeriu2011radon, banjanac2012daily, VICTOR2019105118}]. These reported patterns are similar to that of $\gamma$-ray flux. Usually, the radon gas is exhaled from the ground, and its intensity depends on the geological and soil properties of the observation site. It is generally considered that the diurnal variation in radon concentration near the ground occurs due to the difference in the ABL or mixing height during day and night. Because of atmospheric heating and cooling, contraction and expansion of the air envelope occur. Being in the gaseous form, transportation of the radon gas in the air after its exhalation from the soil is quite plausible. On fair weather days, once exhaled from the soil, depending on the atmospheric temperature and mixing height, radon can reach up to greater heights (day times) or can get accumulated in the lower heights (night times). Under non-turbulent conditions, during night-times as the temperature of the surface-adjacent atmospheric layer is cooled down, the exhaled radon does not get transported to the higher heights, and an inversion layer is formed keeping the radon accumulated near the surface. When the sun rises again, the temperature starts increasing; thus, a thin but stable boundary layer formed due to cooling gets perturbed. As a result, even though the exhalation from soil continues, the concentration detected at different heights during different times differ depending on the mixing height or ABL \citep{galeriu2011radon}. This pattern is well-defined during sunny days or warm months in general \citep{el2001diurnal,desideri2006monitoring}, as the solar heating-induced vertical mixing is more efficient in the warmer season.  

Another approach is based on the fact that the temperature difference between soil and air is different during the day and night, which can result in the difference in the exhalation of radon from the ground. \citet{schubert2002diurnal} reported that during night-time, the temperature of the soil is higher than the temperature of the air above; hence, there is a more pronounced upward-directed convective flux of radon. This increases the overall diffusive transport, which results in the maximum concentration in the morning. When the situation is reversed, a downward convective flux emerges, reducing the overall upward diffusive radon transport. This also explains why there is a decline in $\gamma$ flux during the daytime \citep{schubert2002diurnal}. However, \citet{gogolak1977variation} have pointed out that the diurnal variations in the exhalation rate have a small effect on concentration profiles compared to that of changes in the vertical diffusivity.

Recently, \citet{VICTOR2019105118} presented results on diurnal and seasonal variations of $^{222}$Rn using a radon detector (RTM 2200) at Pune, India. They have explained the results in terms of surface meteorology, soil moisture, soil temperature, vertical wind velocity, local topography, and development of ABL. They report that $^{222}$Rn concentration increased with soil moisture content of $<$ 16\%, rapidly dropped down for dry soil values and then remained almost constant with increasing soil moisture \citep{VICTOR2019105118}; while \citet{hosoda2007effect} had showed experimentally that radon exhalation rates increased with increase in moisture content up to 8\% but then showed a decreasing tendency for soil moisture $>$ 8\%. According to \citet{schery1989desorption}, when temperature increases or moisture content rises due to rainfall in a very dry, clay-rich soils, more radon will be exhaled. 

\begin{figure*}
\vspace*{2mm}
\begin{center}
\includegraphics[width=15cm]{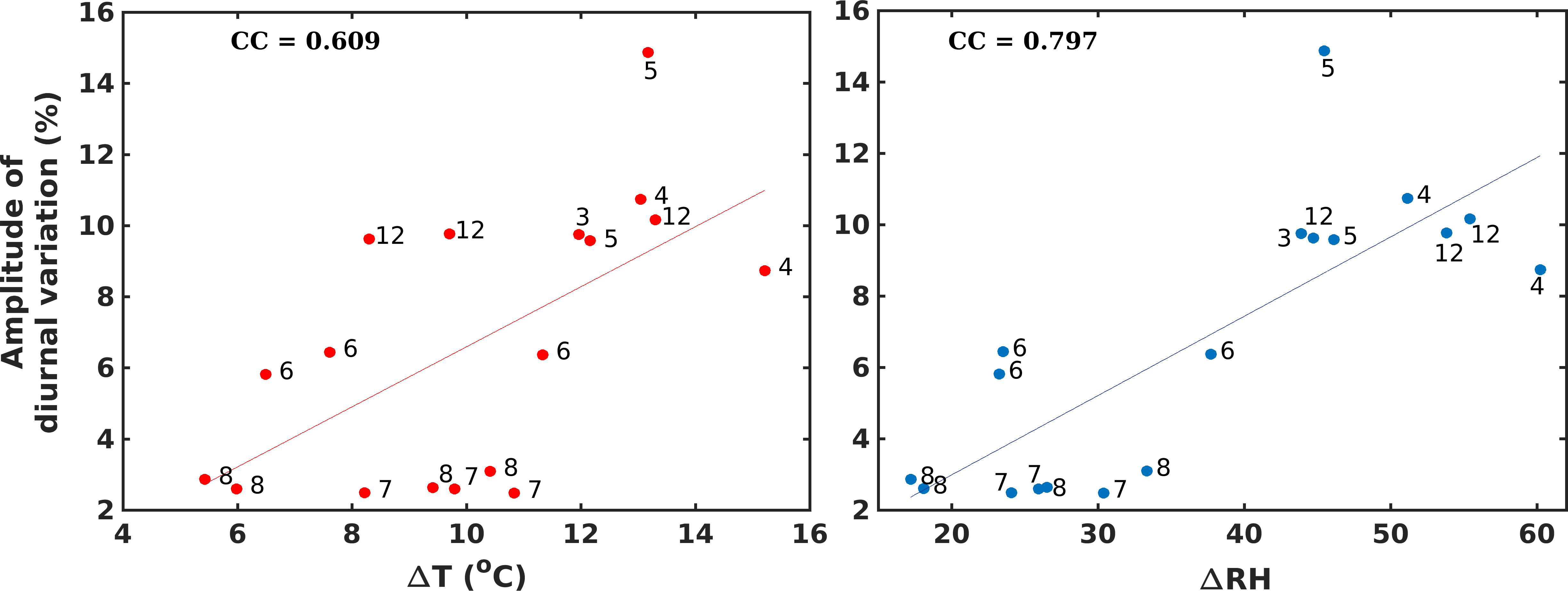}
\end{center}
\caption{\textit{Scatter plots between the diurnal amplitudes of $\gamma$ flux and temperature variations (left) and relative humidity (RH)(right) for individual days (labels show the corresponding month).}}
 \label{fig:ten}
\end{figure*}

\begin{figure*}
\vspace*{2mm}
\begin{center}
\includegraphics[width=14cm]{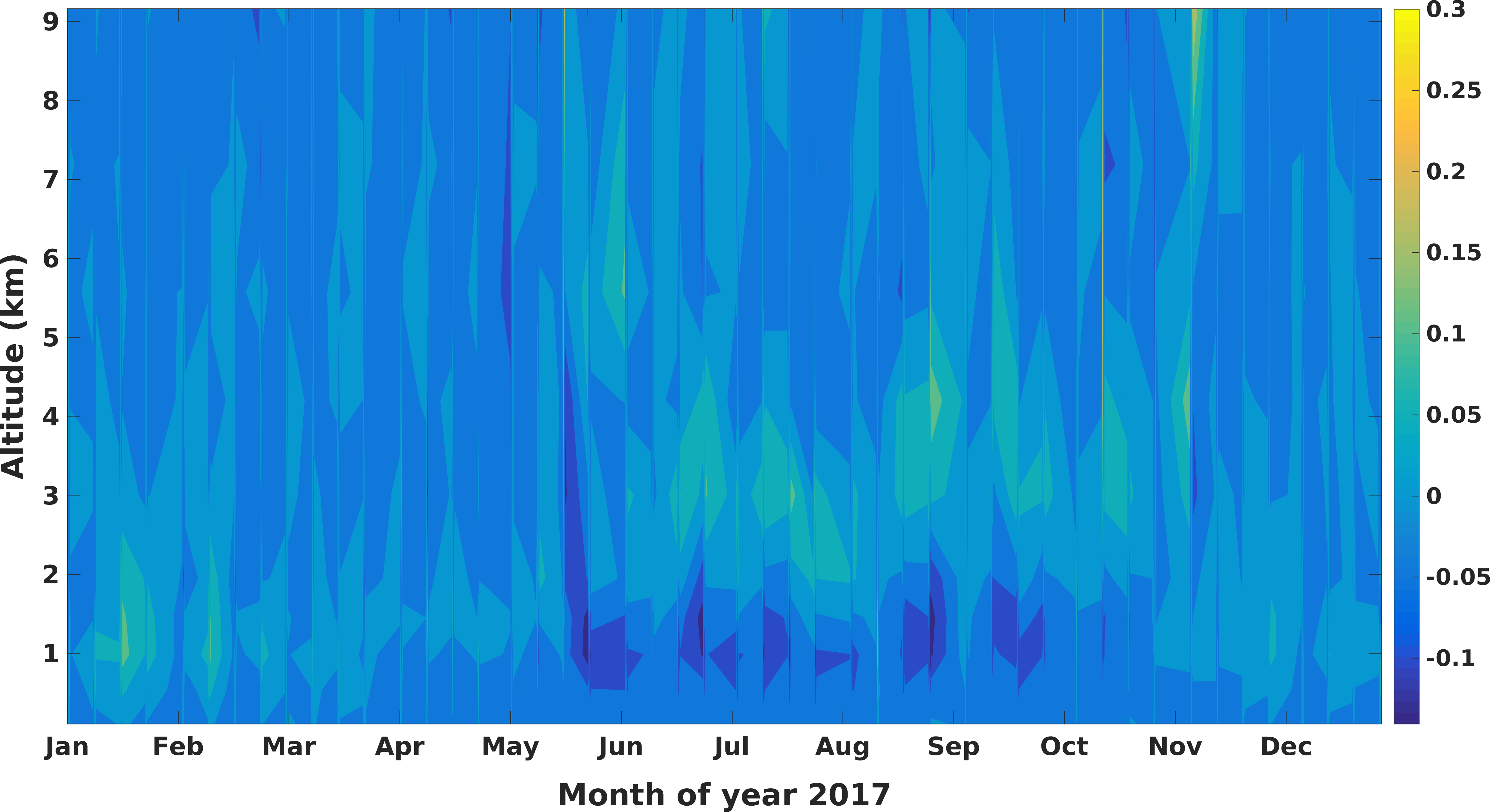}
\end{center}
\caption{\textit{Vertical velocity (m/s) from January to December 2017 ('Contains modified Copernicus Atmosphere Monitoring Service Information [2017]').}}
 \label{fig:eleven}
\end{figure*}

In the present study conducted during January-December 2017, a strong seasonal dependence of the diurnal amplitude (indicated by the daily range of variation) is observed. In the months between June-October, the diurnal amplitude is very weak, although the pattern is still present. The seasonal variation of the diurnal variation is visualised in Figure \ref{fig:five}. The seasonal variation of the two photopeaks of $^{214}$Bi and $^{40}$K also follow a similar trend, whereas $^{208}$Tl shows very weak seasonal variation. Thus the present analysis indicates that the diurnal pattern observed in $\gamma$ flux is essentially due to $^{214}$Bi radionuclide present in the background radioactivity and some contribution from $^{40}$K and $^{208}$Tl. Referring to the decay chains mentioned in the Appendix, observed two photopeaks of $^{214}$Bi are the daughter radionuclide of $^{222}$Rn. Also, $^{208}$Tl is a daughter radionuclide of another radon isotope, $^{220}$Rn, known as {\lq{thoron gas}'}. The radiation dose from $^{220}$Rn and its daughter products is much smaller than $^{222}$Rn and its daughters \citep{united2000sources}; this could also be because of the short half-life of $^{220}$Rn (55 s, against 3.8 days of $^{222}$Rn). Thus, the weak variation observed in $^{208}$Tl can be attributed to the lesser concentration of $^{220}$Rn than $^{222}$Rn. However, $^{40}$K does not fall in the decay chain. Here, one should note that there exists some increase in the counts of the energies lower than the actual photopeak due to the Compton effect. Likewise, associated with the photopeak of $^{214}$Bi (1.764 MeV), there is a major Compton scattering component present in the lower energies coinciding with the $^{40}$K (1.461 MeV) photopeak, which can affect the counts under the photopeak of $^{40}$K. Thus, the pattern reflected in the counts under $^{40}$K (1.461 MeV) could be actually the Compton component of $^{214}$Bi (1.764 MeV). 

Thus, the present study has demonstrated that the diurnal pattern in the $\gamma$-rays is because of the isotopes of radon ($^{220}$Rn, $^{222}$Rn), and their daughter radionuclides. It is not only $^{222}$Rn but also $^{220}$Rn that contributes to the diurnal variations, in spite of its weak and short-living nature. \citet{el2001diurnal}) had studied the radon progeny concentration in the atmosphere for diurnal and seasonal changes using one-year data; however, they had used $\alpha$-spectrometry. Although they conclude the same about diurnal variation, they suspect that the observed considerable seasonal variations are due to meteorological factors. They also found that in cloudy and windy weather, the difference between the short-lived radon progeny concentration during the day and night was small. Their remark about wind is that turbulence is increased by a strong wind, which tends to distribute radon progeny over a large area and to a great vertical depth in the atmospheric air. However, this does not explain the low variation we observed on days when the wind is not exceeding 5 m s$^{-1}$ (viz. days from June to October). 

To understand the effect of meteorological parameters on the amplitude of $\gamma$ flux, it will be interesting to see the range of temperature and humidity against the amplitude of $\gamma$. In Figure \ref{fig:ten}, range of $\gamma$ flux variation ($\Delta$), in \% and the maximum difference in temperature variation in a day ($\Delta{T}$), in $^o$C as well as relative humidity (RH) is compared for individual days. The days are selected from the list in Table \ref{table:days} as per the availability of temperature data, either from in-house AWS or IMD AWS. Additionally, three more days are selected apart from the list with 5 $<\Delta{T}<$ 7 $^o$C, to include the lower range of temperature below 7 $^o$C. A least square linear fit is shown by the solid line. Diurnal amplitude and $\Delta {T}$ do show a good correlation with the Pearson correlation coefficient (CC) of 0.609. The data points which are out of the general linear trend are from the months June to October (indicated by the labels) and have very low diurnal amplitude. For the relative humidity (Figure \ref{fig:ten}b), the CC value is 0.797. 

\citet{VICTOR2019105118} observed that the diurnal variations of radon concentration during winter, post-monsoon, and pre-monsoon seasons were similar in shape but with decreasing amplitudes. However, during the monsoon season, radon concentration was the lowest and remained almost constant throughout the day \citep{VICTOR2019105118}. They explained their results based on the variations in soil moisture due to the monsoon and capping effect of precipitation. In our case as well, it is clearly evident from the observations in Figures \ref{fig:four} and \ref{fig:five} that the percentage change is minimum from June to October. However, the observation site in the present work receives  rainfall  normally  in  the months  of  October  and  November  during  the  northeast monsoon, while occasional rains occur from June to September--when the southwest monsoon prevails in the Indian subcontinent \citep{panneerselvam2007diurnal}. Thus the main monsoon months at the observation sites of \citet{VICTOR2019105118} and ours are different, and yet the lower amplitudes of radon and its daughter nuclides occur during the same months, which is intriguing. The explanation for the present observed results is sought in terms of the winds, as presented below. 

Moderate-to-intense southwesterly winds dominate during one half (April–September) of the year over the observation site, and moderate north-easterlies during the other half (October–March) of the year \citep{panneerselvam2007diurnal}. From the vertical wind velocity profile presented in Figure \ref{fig:eleven}, from the end of May to September, negative values, i.e., downward winds prevail at altitudes 0.5 km -- 2 km. This suppresses the vertically upward propagation of radon, and the radon exhaled from the soil remains in this thin layer of the atmosphere. This would result in the lower amplitude of diurnal variation, as radon would remain in a thin layer in the lower atmosphere during the day as well as night. The mixing height for radon would not change from day-time to night-time as compared to the months when the vertical wind was positive, i.e., upwards, and mixing heights would vary extensively for day and night.  Intermittent rain spells can increase the soil moisture releasing radon until it reaches a certain limit (as specified by \citet{hosoda2007effect, VICTOR2019105118}). After a spell of rainfall, $\gamma$-ray flux can be enhanced due to rainout and washout processes \citep{takeuchi1982rainout}; however, those are short term enhancements and are not considered in the diurnal amplitudes.  

As mentioned earlier, the observation site receives rainfall mainly due to northeastern monsoon, while southwest monsoon is not experienced significantly. However, the neighbouring state Kerala experiences a heavy (southwest and northeast) monsoon. The orographic feature of the Western Ghats creates such a situation. Particularly, we can consider the case of Agasthyamalai hills situated at $\sim$ 45 km away (West) from the observation site. Although the tallest peak in Agasthyamalai hills is $\sim$ 1800 m; generally, these hills have an elevation of around 1000 m or even less. Even though the southwest monsoon does not cross these hills to much extent, strong horizontal winds can travel across the mountain range, affecting the atmosphere on the other side (i.e., Tamil Nadu). The intense southeasterly horizontal winds can affect the vertical motion of air parcels in an air column, consequently producing downward dominated vertical winds. This would explain why wind observations at Tirunelveli, Tamil Nadu are similar to those made from Pune, Maharashtra. Even though the observation site does not experience the southwest monsoon, the winds carrying them affect the atmospheric column over the observation site, resulting in the downward vertical wind, which further suppresses the exhaled radon. Thus, during the southwest monsoon period (June to October), radon movement in the air column is restricted in a thin layer, giving smaller diurnal variation amplitudes in total $\gamma$-ray flux.

\section{Conclusion}
The present work carried out to understand the diurnal variation of the $\gamma$-ray spectrum and its causes recorded by NaI (Tl) scintillation detector at Tirunelveli, South India concludes the following: 
\begin{itemize}
\item There exists a distinct and significant ( $>$ 10\%) diurnal pattern in the total number of $\gamma$-ray counts. 
\item The counts start decreasing after sunrise and show gradual recovery after sunset. 
\item The diurnal pattern is present only in the energies related to the terrestrial background radioactivity (500 keV -- 2.7 MeV). 
\item The study demonstrates that the pattern is associated with the radioactivity of isotopes of radon and their daughter radionuclides. \item The diurnal variation is predominantly due to $^{222}$Rn and its daughter radionuclides. Thoron gas i.e., $^{220}$Rn, though in small proportion, also contributes to the observed diurnal pattern via its daughter product $^{208}$Tl.
\item The amplitude of the diurnal variation is found to have seasonal dependence, with the lowest amplitude during June-October ($\sim$ 2\%), and the highest in April-May months ($\sim$ 14\%). \item Good correlation of the amplitude of diurnal variation with the range of temperature variation (CC = 0.609) as well with the relative humidity (CC = 0.797) is observed.
\item The pattern arises due to changing contribution from radioactivity of radon progenies due to generation of the inversion layer. The distribution of concentration of radon and thoron gases in an air column changes with the vertical mixing and ABL.
\item $\gamma$-rays with energies $>$ 2.7 MeV resulting from the cosmic ray interactions with the atmospheric particles do not exhibit any detectable diurnal pattern. 
\end{itemize}

\section*{Data Availability}
The $\gamma$-ray data from NaI(Tl) detector used for the current study is available on request through proper permissions from the Director, Indian Institute of Geomagnetism (IIG). The vertical wind dataset is available at [\textit{https://cds.climate.copernicus.eu/cdsapp\#!/dataset/reanalysis-era5-pressure-levels}] under public domain.

\section*{Acknowledgements}

The experimental set up at Tirunelveli is operated by the Indian Institute of Geomagnetism. This work is supported by the Department of Science and Technology, Government of India. Authors are grateful to corresponding authorities of IMD and ECMWF for providing respective datasets. We acknowledge that neither the European Commission nor ECMWF is responsible for any use that may be made of the Copernicus Information or Data this publication contains. 

AB acknowledges the support by the NASA Living With a Star Jack Eddy Postdoctoral Fellowship Program, administered by UCARs Cooperative Programs for the Advancement of Earth System Science (CPAESS).

\section*{Author contributions statement}

All authors participated in designing the experiment. GD carried out the data analysis. GD, GV, AB, and AR contributed to the analysis of results, discussion, and writing the manuscript.

\section*{Additional information}
\textbf{Competing interests}: The authors declare no competing interests.
%\clearpage
\newpage
\bibliography{references}
\newpage
\appendix
\textbf{Decay chains of radioactive isotopes of Thorium and Uranium ($^{232}$Th, $^{235}$U, and $^{238}$U) }
\begin{figure*}
\vspace*{2mm}
\begin{center}
\includegraphics[width=12cm]{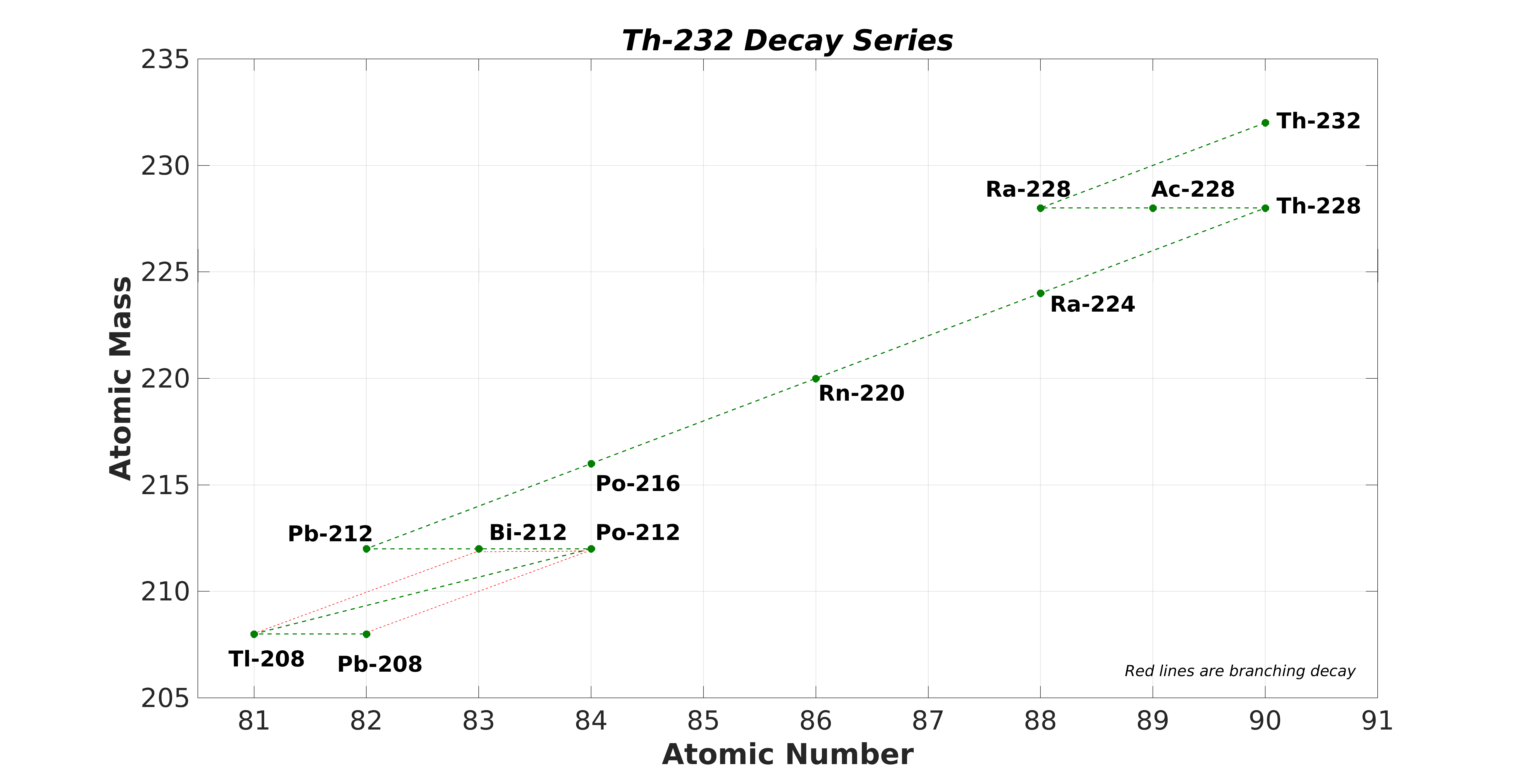} 
\end{center}
 \label{fig:th232}
\end{figure*}

\begin{figure*}
\vspace*{2mm}
\begin{center}
\includegraphics[width=12cm]{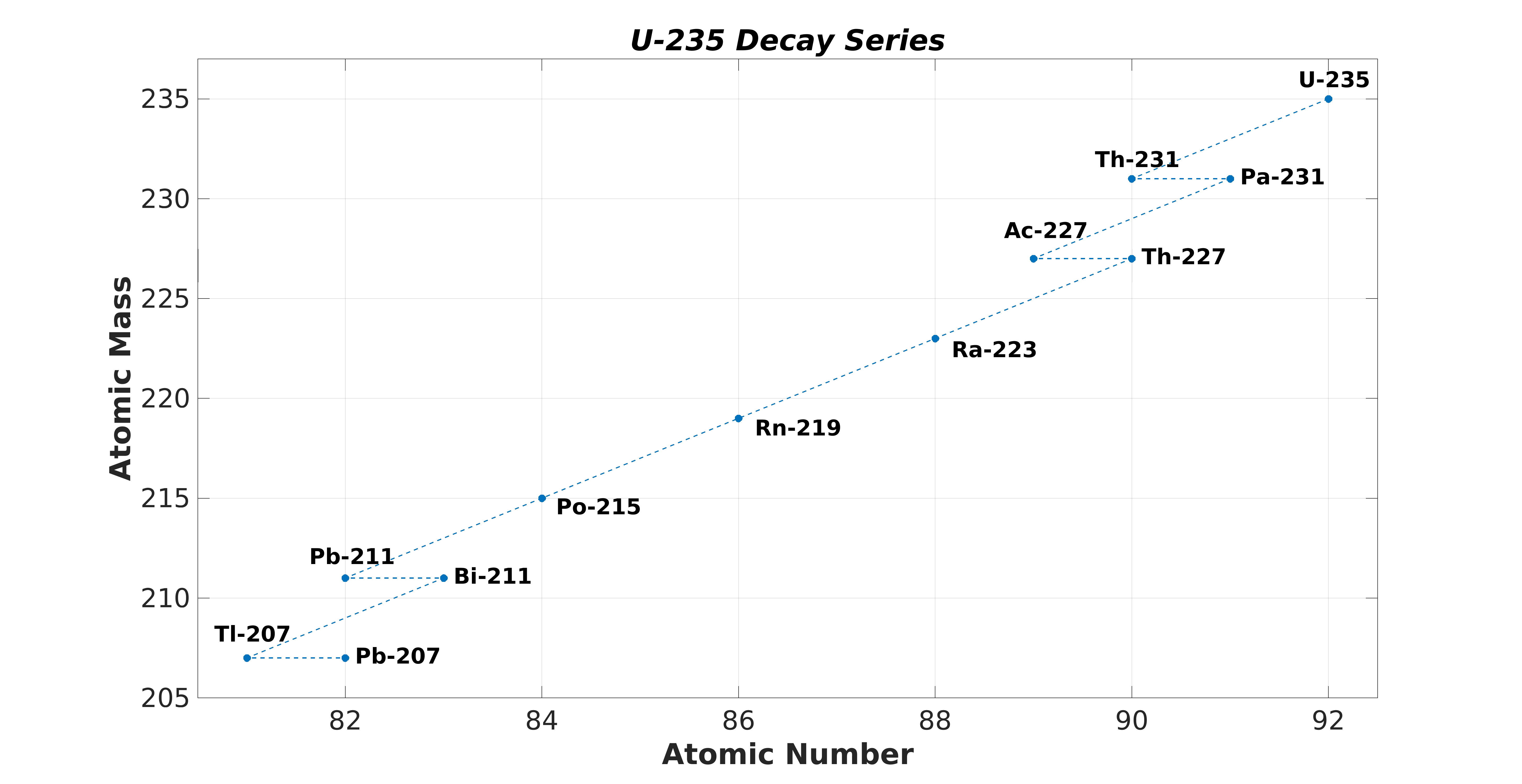} 
\end{center}
 \label{fig:u235}
\end{figure*}

\begin{figure*}
\vspace*{2mm}
\begin{center}
\includegraphics[width=12cm]{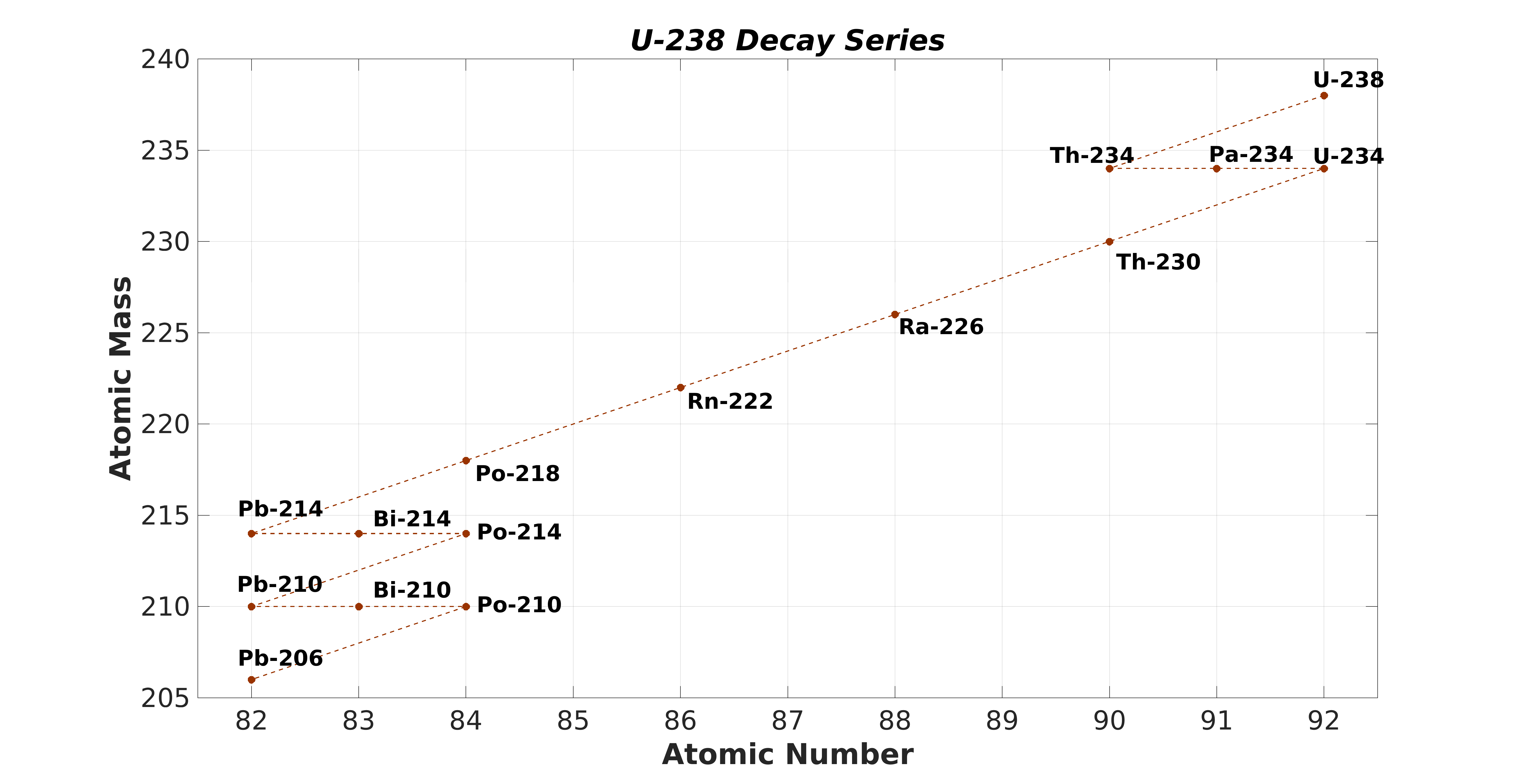} 
\end{center}
 \label{fig:u238}
\end{figure*}

\end{document}